\documentclass{article}
\usepackage[top=1.5in, bottom=1.5in, left=1in, right=1in]{geometry}
\usepackage{graphicx}
\usepackage{multirow}
\usepackage{subcaption}
\usepackage{psfrag}
\usepackage{rotating}
\usepackage{amsmath}
\usepackage{algpseudocode}
\usepackage{tkz-graph}
\usetikzlibrary{arrows}
\usepackage{xcolor,colortbl}
\usepackage[T1]{fontenc}
\usepackage[T1]{fontenc}
\usepackage[utf8]{inputenc}
\usepackage{natbib}
\usepackage{pifont} 
\usepackage{amsmath,amssymb}
\usepackage{authblk}
\usepackage{natbib}
\usepackage{bbm}
\usepackage{bm}
\usepackage{hyperref}
\usepackage{cleveref}
\DeclareMathOperator*{\argmax}{arg\,max}

\DeclareUnicodeCharacter{2212}{-}
\numberwithin{equation}{section}

\title{Model-based inference of conditional extreme value distributions with hydrological applications}
\author[1]{Ross Towe\thanks{r.towe@lancaster.ac.uk}}
\author[2]{Jonathan Tawn}
\author[3,4]{Rob Lamb}
\author[2]{Chris Sherlock}
\affil[1]{School of Computing and Communications, Lancaster University, United Kingdom}
\affil[2]{Department of Mathematics and Statistics, Lancaster University, United Kingdom}
\affil[3]{JBA Trust, Skipton, United Kingdom}
\affil[4]{Lancaster Environment Centre, Lancaster University, United Kingdom}
\begin{document}
\maketitle
\begin{abstract}
Multivariate extreme value models are used to estimate joint risk in a number of applications, with a particular focus on environmental fields ranging from climatology and hydrology to oceanography and seismic hazards. The semi-parametric conditional extreme value model of \citet{HefTawn2004} involving a multivariate regression provides the most suitable of current statistical models in terms of its flexibility to handle a range of extremal dependence classes. 
However, the standard inference for the joint distribution of the residuals of this model is suffers from the curse of dimensionality since in a $d$-dimensional application it involves a $d-1$-dimensional non-parametric density estimator, which requires, for accuracy, a number points and commensurate effort that is exponential in $d$. Furthermore, it does not allow for any partially missing observations to be included and a previous proposal to address this is extremely computationally intensive, making its use prohibitive if the proportion of missing data is non-trivial. 
We propose to replace the $d-1$-dimensional non-parametric density estimator with a model-based copula with univariate marginal densities estimated using kernel methods. This approach provides statistically and computationally efficient estimates whatever the dimension, $d$ or the degree of missing data. Evidence is presented to show that the benefits of this approach substantially outweigh potential mis-specification errors. The methods are illustrated through the analysis of UK river flow data at a network of 46 sites and assessing the rarity of the 2015 floods in north west England.
\end{abstract}

\textit{Keywords: Copula, dependence modelling, missing values, multivariate extreme value theory and spatial flood risk assessment.}

\section{Introduction}\label{sec:Int}

Widespread flooding, such as the events of winter 2015/2016 in the UK, demonstrate the importance of understanding the likelihood of multiple locations experiencing extreme river flows. During these events 43,000 homes were left without power and the estimated damages totalled \pounds 1.3-1.9 billion \citep{EA2018}. For flood risk management and insurance purposes, we are interested in understanding the joint probability of events such as those observed in winter 2015/2016 and the likely nature of events that are even more extreme. 

Let $R_{i}$ represent the river flow at gauge $i$ at a given time with corresponding location $\mathbf{s}_{i}$.
Consider $n$ independent and identically distributed realisations of the variable $\mathbf{R}=\left(R_1,\ldots,R_d\right)$, with this variable representing the joint behaviour of river flows at $d$ gauges recorded over a given time period. From observations of these variables we are interested in estimating marginal and joint probabilities. For example, for assessing the rarity of the 5th December 2015 event in north west England, let $v_i$ be the measured flood value in this event for the $i$th gauge in the region. Then we need to know about marginal risk assessment at gauge $i$, through estimating the probabilities 
$\mathbb{P}\left(R_i>v_i \right)$, $i=1,\ldots ,d$, and for joint risk assessment the probability 
$\mathbb{P}\left(\mathbf{R}\in A \right)$ where $A=\{\mathbf{r}=(r_1, \ldots ,r_d)\in \mathbb{R}^d: r_i>v_i, i=1,\ldots ,d\}$. More generally 
we are interested in estimating the probability $\mathbb{P}\left(\mathbf{R}\in A \right)$ where the set $A\subset \mathbb{R}^d$ is extreme for at least one component, $R_i$ of $\mathbf{R}$ say, so that for all $\mathbf{r}\in A$, then $r_i>q_i$ with $q_i$ a high quantile for variable $i$. 

For modelling spatial multivariate extremes data, the most widely used approach uses max-stable processes \citep{Davison2012,Asadi2015}. However, max-stable processes imply a strong form of extremal dependence, termed asymptotic dependence, in which the largest values at each site, over different events, can occur in the same particular flood event. The assumption of asymptotic dependence is probably reasonable for local-scale studies, such as in a mesoscale river basin, however, for larger-scale studies, such as widespread studies across regions of the UK, this dependence assumption is highly restrictive as the largest values at different sites are unlikely to be occur in a single event. 

Recent developments in statistical modelling of hydrological extremes allow us to now place such widespread events into a probabilistic framework \citep{KeefH2009,Lamb2010,Keef2013b}. Underpinning such methods is the theory of multivariate conditional extremes of \citet{HefTawn2004}. This approach is able to handle the
required mixture of both asymptotic dependence and asymptotic independence (a weaker form of extremal dependence than asymptotic dependence, see Section~\ref{sec:chi}, for both extremal dependence structures that are identified in river flow data. Their conditional dependence model is formed through a semi-parametric regression with parametric components describing variation in the means and the variances of the joint conditional distribution, and the joint distribution of the multivariate residuals being estimated empirically. The parametric components  determine the core extremal dependence features, such as whether subsets of the variables are asymptotically dependent or asymptotically independent, and model across the range of possible dependence structures.

For hydrological applications the method needs to be able to: handle high dimensions (typically for $10-1,000$ sites); give realistic simulations of multivariate extreme events; enable the estimation of the risk of events which are simultaneously rare at all and/or many sites; and allow covariates to be incorporated. Direct application of the \citet{HefTawn2004} method fails when dealing with any one of these issues, let alone being able to address all of these aspects in one analysis. The key problem with \citet{HefTawn2004} is that the empirical multivariate residual modelling suffers from the curse of dimensionality, which along with its restriction to its reliance on the previously observed residuals, means that extrapolations to rarer events corresponds to relocated and rescaled versions of past events. These events have poor coverage over the extremal regions of the sample space in high-dimensional studies and so lead to inefficient inference.

An additional complication that hydrological applications bring is that of missing data. Here we assume the data to be missing at random. Data are likely to be missing when gauges are installed at different times or gauges become faulty. The \citet{HefTawn2004} approach, with its empirical residual distribution model, can only be applied for a $d$-dimensional problem when all components of the $d$-dimensional variable are observed. One approach would be to only analyse complete vector observations. This approach is highly restrictive, for example when considering the whole of the UK river network, with $\sim 1000$ gauges, which were considered as part of the National Flood Resilience Review \citep{Tawn2018}, no concurrent observations are observed at all locations, and hence leads to highly inefficient inference
about extreme events. An alternative approach, proposed by  \citet{Keef2009}, is to replace these missing data, via infilling all the missing residuals with jointly generated multiple samples for the distribution of missing residuals given the observed residuals. This approach, which assumes a Gaussian copula for the joint distribution of missing and observed residuals only, and treats fully observed variables empirically, is hugely computationally intensive when the amount of missing data is non-trivial. Critically it fails to address all the other problems with the \citet{HefTawn2004} method that are described above.


Instead, in this paper the full residual distribution is modelled semi-parametrically: one-dimensional kernel-smoothed distribution functions capture the marginal behaviours of the observed residuals and a Gaussian copula is used for their dependence structure \citep{Joe2014}. Although this change in approach may at first seem rather small it has major implications for the applicability of the method, in that it addresses all the problematic issues of \citet{HefTawn2004} as well as handling large volumes of missing data efficiently. The primary reasons for its success are that as it the removes the problems of the curse of dimensionality and the choice of copula is flexible and parsimonious.
Of course there is a cost to be incurred by this modelling approach, as there is no theoretical motivation to support this assumption. 
However, here we show plenty of evidence to suggest that the Gaussian copula is suitable for modelling the residual copula structure, mainly as it plays a secondary role in capturing the extremal dependence relative to the \citet{HefTawn2004} regression parameters. It is important though to have strong diagnostic tools to assess departures from this model and a clear understanding of the effects of mis-specification. 
This paper is the first that looks carefully at these aspects and finds that there are substantial improvements from the added flexibility and the more efficient use of the data on the estimation of probabilities of rare events.

The \citet{HefTawn2004} model is explained briefly in Section~\ref{sec:ReHT}; with the extensions that we propose and their connections with previously adopted Gaussianity assumptions given in Section~\ref{sec:SimHT}. The methodology for testing the validity of our proposed approach, including dealing with missing data, is detailed in Section~\ref{sec:TeGa}. The comparisons with existing approaches to handle missing values are presented in Section~\ref{sec:MisCEVM}. A generic simulation algorithm for the proposed conditional extreme value model and techniques for estimating probabilities of extreme joint events are given in Section~\ref{sec:SimAg}.
Then, examples of the proposed methodology are given in Sections~\ref{sec:SimSt} and \ref{sec:WkEx} for simulated and observed data respectively. The methodology is applied to study widespread flooding in north west England, the success of the different methods is compared through estimated probabilities of joint flood risk.  The paper finishes with a discussion which considers ways in which the model can be made more parsimonious. Throughout the paper all vector algebra is to be interpreted as being componentwise.

\section{The Heffernan and Tawn model}\label{sec:ReHT}
\subsection{Marginal model}\label{sec:MarTr}

The model for the marginal distributions of $\mathbf{R}$ has two components, separated using the predetermined threshold level $u_i$ for variable $R_i~(i=1, \ldots ,d)$. For a univariate random variable $R_i$, asymptotic theory considers the distribution of excesses over a threshold of $u_{i}$, scaled by some function $c(u_{i})>0$, i.e., $\mathbb{P}\left(c(u_{i})(R_{i}-u_{i})\geq r | R_{i}>u_{i}\right)$, with $r>0$; if this converges to a non-degenerate limit as $u_{i}$ tends to the upper endpoint of the distribution of $R_{i}$ then the limit distribution can only be the generalised Pareto distribution \citep{Pickands1971}. If it is assumed that this limit model holds exactly for some large enough threshold $u_{i}$ it follows that 
\vspace{-0.8cm}
\begin{center}
\begin{equation}
\mathbb{P}\left(R_{i}\geq r | R_{i}>u_{i}\right)=\left[1+\xi_{i}(r-u_{i})/\sigma_{i}\right]^{-1/\xi_{i}}_{+},~\mbox{for~}r~>u_{i},
\label{eq:GP}
\end{equation}
\end{center}
with the scale parameter $\sigma_i>0$ and the shape parameter $\xi_i\in\mathbb{R}$ and the notation $[r]_{+}=\max\left(r,0\right)$  \citep{Davison1990}. Above the threshold, the generalised Pareto distribution (GPD) is adopted. 
For those points below the threshold $u_i$, there is no theoretical justification for any particular model choice, so instead a kernel smoothed empirical cumulative distribution function
$\tilde{F}_i(r)$ of $R_i$ is used.  Thus 
\vspace{-0.5cm}
\begin{center}
\begin{equation}
F_{i}(r) = \begin{cases} \tilde{F}_{i}(r) & \mbox{for } r\leq u_{i},\\ 
								1-\phi_{u_{i}}\left[1+\xi_{i}(r-u_{i})/\sigma_{i}\right]^{-1/\xi_{i}}_{+} & \mbox{for } r>u_{i},\end{cases}		
\label{eq:CDFDef}
\end{equation}
\end{center}
where $\phi_{u_i}=1-\tilde{F}_i(u_i)$ is the probability of an exceedance above the threshold $u_i$. 

Estimating $(\sigma_{i},\xi_{i})$ for each gauge separately can lead to inefficient inference as the spatial coherence and dependence of $R_{i}$ over gauges suggests that $(\sigma_{i},\xi_{i})$ and $(\sigma_j,\xi_j)$ should be more similar when gauges $i$ and $j$ are closer together. Methods such as the covariate hierarchical/latent variable models that spatially smooth the GPD parameters have been developed by  \citet{Cooley2007} and \citet{Cooley2010}. These models are ideal in the generation of marginal quantile maps as they share information from neighbouring sites to reduce any uncertainty in the estimation of quantiles. As the focus of this paper is on dependence modelling we restrict ourselves to separate marginal fits, but recognise this typically can be improved upon.

To help estimate the dependence structure of the random variable $\mathbf{R}$, the data are transformed componentwise to a variable $\mathbf{Y}=(Y_{1},\ldots,Y_{d})$, with common Laplace margins, via the transform
\vspace{-1cm}
\begin{center}
\begin{equation}
Y_i = \begin{cases} \log\left\{2F_{i}\left(R_{i}\right)\right\} & \mbox{for } F_i\left(R_{i}\right)<0.5,\\ 
								\log\left\{2\left[1-F_i\left(R_{i}\right)\right]\right\} & \mbox{for } F_i\left(R_{i}\right)\geq 0.5,\end{cases}		
\label{eq:LapTr}
\end{equation}
\end{center}
for $i=1,\ldots,d$ and where $F_{i}$ is given in equation \eqref{eq:CDFDef}. The transformation to Laplace margins means that $\mathbb{P}\left(Y_i>y+v\vert Y_i>v\right)=\mathbb{P}\left(Y_i < -(y+v)\vert Y_i < -v\right)=\exp\left(-y\right)$  for $y>0$, and $v>0$. Therefore, the marginal random variables of $\mathbf{Y}$ now have exponential upper and lower tails. This is a minor deviation from the \citet{HefTawn2004} approach, as they transform to Gumbel margins, but the use of Laplace margins unifies the handling of positive and negative dependence \citep{Keef2013a}. 

\subsection{Introduction to Extremal Dependence Properties}
\label{sec:chi}
Extremal dependence properties need to be studied for all combinations of the variables as, unlike for multivariate Gaussian 
distribution, not all dependence is determined by the set of pairwise dependences. So consider $C\in 2^D$ with $|C|\ge 2$ and $D=(1,\ldots ,d)$, then define a measure of extremal dependence for variables $\{R_i; i\in C\}$ by 
\vspace{-0.8cm}
\begin{center}
\begin{equation}
\chi_{C}=\lim_{p\rightarrow 1}\mathbb{P}\left(F_{i}(R_{i})>p,~i\in C \right)/(1-p)
=\lim_{v\rightarrow \infty}\mathbb{P}\left(Y_{i}>v,~i\in C \right)2\exp(v),\nonumber
\end{equation}
\end{center}
where $F_{i}$ is the marginal distribution function of $R_{i}$. If $\chi_{C}>0~(\chi_{C}=0)$ the variables in $C$ are jointly asymptotically dependent (asymptotically independent). Here $\chi_{C}>0$ means that extreme events can occur simultaneously over all sites in $C$, whereas if $\chi_{C}=0$ such events are impossible for the set of sites $C$. Clearly for $B\subset C$,  it is possible that $\chi_{C}=0$ and $\chi_{B}>0$ but if $\chi_{B}=0$ then $\chi_{C}=0$. Thus it is possible to have asymptotic dependence locally but asymptotic independence over all sites. 

If a copula model is used the extremal dependence structure is pre-determined by the choice of the copula before the model is fitted. 
For example the class of bivariate extreme value distribution copulas have $\chi_{1,2}>0$ (unless the variables are independent)  and the class of multivariate Gaussian copula, with parameters $\{\rho_{i,j}; i \not= j\in D\}$, have $\chi_{C}=0$ (unless $\rho_{ij}=1$ for all $i,j\in C$ for all $C\in 2^D$ with $|C|\ge 2$). Other standard  copula models typically can only handle one of the two classes of extremal dependence \citep{Heffernan2000}.
As both of the extremal dependence classes are typically observed in extreme river flow data sets, see \citet{Keef2009,Tawn2018}, a standard copula approach is almost never sufficiently flexible. Instead, like with univariate extremes, we appeal to asymptotic formulations to motivate a class of models specific to the tail region. These models allow any possible combination of feasible $\chi_{C}$ values for $C\in 2^D$.

\subsection{Extremal Model for Conditional Dependence}\label{sec:ConDep}
After making the transformation given in equation \eqref{eq:LapTr}, the extremal behaviour of the joint tail of the random variable $\mathbf{Y}$ can now be determined. The approach models $\mathbf{Y}$ given that at least one of its elements is extreme, i.e., given that $\max(\mathbf{Y})>v$ for large $v$, where $v$ is a dependence threshold. 

First assume that $Y_{1}>v$, then the joint distribution of the $(d-1)$ remaining variables $\mathbf{Y}_{-1}=(Y_2,\ldots,Y_d)$ is modelled conditional on $Y_1$ being above $v$. The approach is motivated by the following asymptotic formulation studied by \citet{HefTawn2004} and \citet{Heffernan2007}. The underlying idea is to see how $\mathbf{Y}_{-1}$ behaves as $Y_1$ gets large. In order to avoid non-degeneracy of the limiting conditional distribution of $\mathbf{Y}_{-1}$ as $Y_1$ tends to its upper end point it is sensible to look for a componentwise location-scale transformation of $\mathbf{Y}_{-1}$ using functions of $Y_1$. As dependence between $Y_1$ and each component of $\mathbf{Y}_{-1}$ may be different these location-scale transformations need to have the flexibility to be different for each component. This leads to the assumption that there exists normalising functions, $\mathbf{a}(.):\mathbb{R}\rightarrow \mathbb{R}^{d-1}$ and $\mathbf{b}(.)>\boldsymbol{0}:\mathbb{R}\rightarrow \mathbb{R}^{d-1}_{+}$ such that the following limit probability holds for $y>0$
\vspace{-0.8cm}
\begin{center}
\begin{equation}
\lim_{v\rightarrow \infty}\mathbb{P}\left(\frac{\mathbf{Y}_{-1}-\mathbf{a}(Y_1)}{\mathbf{b}
(Y_1)}\leq \mathbf{z},~Y_1-v>y ~\rvert ~Y_1>v\right)=\exp\left(-y\right)G(\mathbf{z})
\label{eq:LimPro}
\end{equation}
\end{center}
where the joint distribution function $G(\mathbf{z})$ is non-degenerate in each margin and has no mass for any margin at infinity. The first term in the limit given in equation~\eqref{eq:LimPro} arises from the fact that $Y_1$  follows a standard Laplace distribution. The second term in the limit characterises the behaviour of $\mathbf{Y}_{-1}|Y_1>v$ in terms of the limiting distribution function $G(\mathbf{z})$ along with the location $\mathbf{a}(.)$ and scale $\mathbf{b}(.)$ functions. It is assumed that the normalisations of the variables $\mathbf{Y}_{-1}$ and $Y_1$ are independent in the limit. This last assumption parallels that in classical point process models for multivariate extremes and regularly varying distributions \citep{ColesTawn1991, Resnick2013}, with radial and angular representations being assumed to be independent in the limit as the radial variable tends to infinity. \citet{HefTawn2004} show that  formulation~\eqref{eq:LimPro} holds for all standard copula models.

As a result of equation~\eqref{eq:LimPro}, $G(\mathbf{z})$ is the limiting conditional distribution of
\vspace{-0.8cm}
\begin{center}
\begin{equation}
\mathbf{Z}=\frac{\mathbf{Y}_{-1}-\mathbf{a}(Y_1)}{\mathbf{b}(Y_1)},~\mbox{given~}Y_1 > v~\mbox{as~}v\rightarrow\infty,
\label{eq:ResDef}
\end{equation}
\end{center}
where $\mathbf{Z}\sim G$ and we call $\mathbf{Z}$ the residual of the conditional extreme value model. The result of the limits given in equations~\eqref{eq:LimPro} and~\eqref{eq:ResDef} is that $\mathbf{Z}$ and $Y_1$ are independent given that $Y_1>v$ in the limit as $v\rightarrow\infty$. Similar limits, with potentially different $\mathbf{a}(.)$, $\mathbf{b}(.)$ and $G$ holds for $\mathbf{Y}_{-j} |Y_j>v$ for any $j=2,\ldots,d$. Joining together these $d$ different conditionals we have a model for the joint tail behaviour of $\mathbf{Y}$, when at least one component is large. 

Under weak assumptions on the joint distribution of $\mathbf{Y}$, \citet{Heffernan2007} show that componentwise $\mathbf{a}(\cdot)$ and $\mathbf{b}(\cdot)$ must be regularly varying functions satisfying certain constraints, which for Laplace margins corresponds to each of the components of $\mathbf{a}$ (respectively $\mathbf{b}$) being regularly varying functions of index 1 (respectively less than 1). 
\citet{HefTawn2004}, \citet{Keef2013a} and \citet{Papastathopoulos2016} found that although different classes of extremal dependence have different forms for $\mathbf{a}(.)$ and $\mathbf{b}(.)$, they all can be well approximated in a simple parametric form, which is the dominant power term of the regularly varying functions, i.e., excluding the slowly varying function.
For Laplace margins, this form simplifies to 
\begin{equation}
\mathbf{a}(y)=\boldsymbol{\alpha}y \mbox{ and }\mathbf{b}(y)=y^{\boldsymbol{\beta}},~-\boldsymbol{1}\leq\boldsymbol{\alpha}\leq\boldsymbol{1} \mbox{ and }-\boldsymbol{\infty}<\boldsymbol{\beta}<\boldsymbol{1}
\label{eq:paraAB}
\end{equation}
with $\boldsymbol{\alpha}=\left(\alpha_{2},\ldots,\alpha_{d}\right)$ and $\boldsymbol{\beta}=\left(\beta_{2},\ldots,\beta_{d}\right)$. When $(\alpha_{i},\beta_{i})=(1,0)$ for all $i\in C_{-1}\subset D\backslash \{1\}$ then if $C=C_{-1} \cup \{1\}$
it follows that $\chi_{C}>0$ and the variables indexed by $C$ are asymptotically dependent. Similarly if $\alpha_{i}<1$ for any $i\in C_{-1}$ then $\chi_{C}=0$ and the variables indexed by $C$ are asymptotically independent. Thus $\boldsymbol{\alpha}$ controls the collections of variables which are asymptotically dependent with variable $Y_1$.
It is clear therefore that this model captures all the possible sets of asymptotically independent and dependent variables as set out in Section~\ref{sec:chi}.
This unification of the parametric forms for all dependence classes enables flexible efficient statistical modelling unlike with standard parametric copula modelling. 

\citet{HefTawn2004} assume that limit~\eqref{eq:LimPro} holds exactly above a sufficiently large dependence threshold $v$ and that the normalising functions are given by the parametric forms~\eqref{eq:paraAB}. This leads to the following model:
\vspace{-1cm}
\begin{center}
\begin{equation}
\mathbf{Y}_{-1}=\boldsymbol{\alpha}Y_1+Y_1^{\boldsymbol{\beta}}\mathbf{Z},~\mbox{for~}Y_1>v, 
\label{eq:HTSim}
\end{equation}
\end{center}
where $-\boldsymbol{1}\leq\boldsymbol{\alpha}\leq\boldsymbol{1}$ and $-\boldsymbol{\infty}<\boldsymbol{\beta}<\boldsymbol{1}$ and $\mathbf{Z}\sim G$, where $G$ is a marginally non-degenerate distribution function and the $\mathbf{Z}$ is independent of $Y_1$.  There is no general theoretically justified family of distributions $G$ for the multivariate residuals $\mathbf{Z}$, so \citet{HefTawn2004} assumed that $\mathbf{Z}$ has marginal finite means and variances $\boldsymbol{\mu}$ and $\boldsymbol{\sigma}^{2}$ respectively, where $\boldsymbol{\mu}=\left(\mu_{2},\ldots,\mu_{d}\right)$ and $\boldsymbol{\sigma}=\left(\sigma_{2},\ldots,\sigma_{d}\right)$. As a result, the following expressions for the conditional expectation and variance of $Y_{i}|Y_{1}=y$ can be determined for $y>v$ and $i=2,\ldots,d$,
\vspace{-1cm}
\begin{center}
\begin{eqnarray}
\mathbb{E}[Y_{i} \vert Y_1=y ]&=&\alpha_{i} y+y^{\beta_{i}}\mu_{i},\nonumber\\
\mathbb{V}\mbox{ar}[Y_{i} \vert Y_1=y]&=&(y^{\beta_{i}}\sigma_{i})^{2}.
\label{eq:ExVar}
\end{eqnarray}
\end{center}
\citet{HefTawn2004} model the joint distribution of $\mathbf{Z}$ non-parametrically using an empirical joint distribution, with the specific form of this model presented in Section~\ref{sec:Inf}.

So far we have presented the behaviour of  $\mathbf{Y}|Y_1>v$ for large $v$, or equivalently $\mathbf{Y}|Y_i>v$ for an arbitrary $i\in D$, but we really want the behaviour of $\mathbf{Y}|\max(\mathbf{Y})>v$. This conditional behaviour can be derived from the set of distributions of $\mathbf{Y}|Y_i>v$ for $i\in D$. As the conditioning variable changes to $Y_i$ the norming functions $\mathbf{a}(\cdot)$ and $\mathbf{b}(\cdot)$ as well as the limiting distributions $G$ all change with $i$. We can piece together results from a series of models of the form above. A limitation of this set of models is that self-consistency is not ensured unless specific constraints on these different normalisation and distribution functions are made. A lack of self-consistency may lead to inconsistencies when joint exceedance probabilities are estimated, with the results depending on the choice of conditioning variable. \citet{HefTawn2004} review ways of avoiding this problem with partitioning the sample space and \citet{Liu2014} discuss a number of approaches to reduce this problem. In this paper we will, however, largely look at the individual conditional distributions,  i.e., $\mathbf{Y}|Y_i>v$ for $i\in D$ and not overall joint tail inference.

\subsection{Inference}\label{sec:Inf}

The dependence parameters $\boldsymbol{\alpha}$ and $\boldsymbol{\beta}$ of the \citet{HefTawn2004} model are estimated through pairwise maximum pseudo likelihood for the $n_v$ pairs with $Y_{1}>v$. The pseudo likelihood 
$L\left(\boldsymbol{\alpha},\boldsymbol{\beta},\boldsymbol{\mu},\boldsymbol{\sigma}\right)$ for inference for 
$(\boldsymbol{\alpha},\boldsymbol{\beta})$ is constructed under the temporary working assumption that 
\[
G(\mathbf{z})=\prod_{i=1}^d \Phi\left(\frac{z_i-\mu_i}{\sigma_i}\right),
\]
i.e., independent Gaussian distributions. Hence
\vspace{-0.8cm}
\begin{center}
\begin{equation}
L\left(\boldsymbol{\alpha},\boldsymbol{\beta},\boldsymbol{\mu},\boldsymbol{\sigma}\right)\propto\prod_{i=2}^{d}\prod_{j=1}^{n_v}\frac{1}{y_{ij}
^{\beta_i}\sigma_i}\exp\left\{-\frac{(y_{ij}-\alpha_i y_{1j}-\mu_{i}y_{1j}^{\beta_i})^2}{2(y_{ij}^{\beta_i}\sigma_i)^2}\right\},
\label{eq:NorLike}
\end{equation}
\end{center}
here $-\infty<\mu_i<\infty$, $\sigma_i>0$, $-1\leq\alpha_i\leq1$, and $-\infty<\beta_i<1$, where $y_{ij}$  denotes component $i=1,\ldots,d$ for the $j^{th}$ exceedance of $v$ by $Y_1$. 
The maximum pseudo likelihood estimates $\hat{\boldsymbol{\alpha}}=(\hat{\alpha}_{2},\ldots,\hat{\alpha}_{d})$ and $\hat{\boldsymbol{\beta}}=(\hat{\beta}_{2},\ldots,\hat{\beta}_{d})$ are found, by jointly maximising equation \eqref{eq:NorLike}, with $\boldsymbol{\mu}$ and $\boldsymbol{\sigma}$.

Now we present the \citet{HefTawn2004} modelling and inference for the joint distribution of the residuals. This is where our inference approach outlined in Section~\ref{sec:SimHT} differs. Firstly the 
temporary working assumption of independent Gaussianity of the components of $\mathbf{Z}$ used in the estimation of 
$\boldsymbol{\alpha}$ and $\boldsymbol{\beta}$ is discarded. With the fitted values of these parameters there are $n_v$ observed exceedances of $v$ by $Y_1$, denoted $y_{1j}, j=1, \ldots ,n_v$.
The associated vectors of residuals are $\{\mathbf{z}^{(j)}, j=1, \ldots ,n_v\}$, where $\mathbf{z}^{(j)}=\left(z_{2j},\ldots,z_{dj}\right)$ with its component associated with $Y_i$ given by
 \vspace{-0.7cm}
\begin{center}
\begin{equation}
z_{ij}=\frac{y_{ij}-\hat{\alpha}_{i}y_{1j}}{y_{1j}^{\hat{\beta}_i}},~\mbox{for~}y_{1j}>v,~\mbox{where~}j=1,\ldots,n_v,~i=2,\ldots,d.
\label{eq:RComp}
\end{equation}
\end{center}
\citet{HefTawn2004} estimate the joint distribution function $G$  through the empirical joint distribution function
of these residuals $\mathbf{z}^{(1)},\ldots,\mathbf{z}^{(n_{v})}$.
Extrapolation from the model comes from \eqref{eq:HTSim}, with larger events arising when $Y_{1}$ is larger than the observed events. Due to the independence of $Y_{1}$ and $\mathbf{Z}$, for $Y_{1}>u$, all simulated events are of the form $\mathbf{Y}_{-1}=(\boldsymbol{\alpha}y+y^{\boldsymbol{\beta}}\mathbf{z}^{(j)})$, for $y>v$ and $j=1,\ldots,n_{v}$. This leads to simulated events on Laplace margins being shifted and rescaled versions of past events. Thus the extrapolation is restricted to $n_v$ sets of $1$-dimensional extrapolations, which clearly do not span the required extrapolation space, particularly when $n_v$ is small relative to $d$.

\section{New Modelling Features}\label{sec:SimHT}

\subsection{Semi-parametric inference for \textit{G}}
We model the joint residual distribution $G$ by a semi-parametric joint distribution model with 1-dimensional kernel smoothed marginal distribution functions and a Gaussian copula \citep{Joe2014}. Let $\hat{G}_{i}(z)$ be the kernel smoothed distribution function for observations of $Z_{i}$, then
\vspace{-0.8cm}
\begin{center}
\begin{equation}
\hat{G}_{i}(z)=\frac{1}{n_{v}}\sum_{j=1}^{n_{v}}\Phi\left(\frac{z-z_{ij}}{h_{i}}\right),~\mbox{where~}i=2,\ldots,d
\label{eq:KS}
\end{equation}
\end{center}
with $h_i>0$, the bandwidth \citep{Silverman1986} and $z_{ij}$, given by expression~\eqref{eq:RComp}, corresponding to the $i$th component of the $j$th residual vector when $Y_1>v$. The kernel smoothed distribution provides flexibility as it allows smooth interpolation between observed data points as well as some limited extrapolation and critically it leads to a non-deterministic extrapolation of past events. Our model for the joint distribution function $G$ is then 
\vspace{-0.8cm}
\begin{center}
\begin{equation}
G(\mathbf{z})=\Phi_{d-1}\left(\Phi^{-1}\hat{G}_{i}(z_{i}), i=2,\ldots,d;~\Sigma\right)\label{eq:NorRes}
\end{equation}
\end{center}
where $\mathbf{z}=(z_2, \ldots ,z_d)$, 
$\Phi$ and $\Phi_{d-1}(.,\Sigma)$ are the cumulative distribution functions of a standard univariate Gaussian and a standard $(d-1)$-dimensional Gaussian with correlation matrix $\Sigma$ with $(i,j)$th element $\rho_{ij}$ with $i\neq j=2,\ldots,d$. The use of the componentwise probability integral transformation gives
\[
\mathbf{Z}^{N}=(Z^{N}_{2},\ldots,Z^{N}_{d})=\left\{\Phi^{-1}\left(\hat{G}_{i}(Z_{i})\right),~i=2,\ldots,d\right\}. 
\]
Our copula assumption \eqref{eq:NorRes} then corresponds to $\mathbf{Z}^{N}$ being a $(d-1)$-dimensional standard Gaussian distribution with the correlation matrix $\Sigma$ giving a relationship between the residuals which is fully determined by its bivariate marginals. Furthermore, the Gaussian copula is chosen because it is computationally feasible in high dimensions and is closed to marginalisation and conditioning. The Gaussian copula has an asymptotically independent extremal dependence structure \citep{Ledford1996}, however this property is not restrictive as the joint tails of $\mathbf{Z}$ are not vital for determining the joint tails of $\mathbf{Y}_{-1}|Y_{1}$ as that distribution is a mixture over $Y_1$, for $Y_{1}>v$, so even independent $\mathbf{Z}$ can lead to $\mathbf{Y}_{-1}|Y_{1}>v$ being asymptotically dependent. See Section~\ref{sec:MisCEVM} for details of how to estimate $\Sigma$. 

Unlike the standard \citet{HefTawn2004} approach the residuals are no longer restricted to the sample as the kernel smoothing allows both interpolation and limited extrapolation of the residuals and the Gaussian copula enables new combinations of $\mathbf{Z}$ to occur.

\subsection{Tests of the Gaussian copula assumption}\label{sec:TeGa}
A formal test to check whether the copula it is fairly close to being Gaussian is required to avoid the residual joint model being applied inappropriately. For assessing pairwise dependence, visual inspections of the residual distribution is sometimes sufficient, however this comparison fails to assess the importance of higher-order dependence. In order to assess the full dependence structure, we adopt the methods of \citet{Bortot2000} for assessing Gaussian copula in joint tail regions. 

Consider the set of independent and identically distributed observations of $\mathbf{Z}^N$, which follows a $(d-1)$-dimensional multivariate Gaussian distribution with correlation matrix $\Sigma$. The square of the Mahalanobis distance is defined by 
\vspace{-1cm}
\begin{center}
\begin{equation}
T=\mathbf{Z}^N\Sigma^{-1}\left(\mathbf{Z}^{N}\right)^{'}.
\label{eq:Tstat}
\end{equation}
\end{center}
Then $T$ follows a $\chi_{d-1}^2$ distribution with $\mbox{E}[T]=d-1$ and $\mbox{Var}[T]=2(d-1)$. 
In reality, there are missing (at random) values in the observations of the residual variable $\mathbf{Z}^N$ and the percentage of missing values is not consistent across locations. Therefore, the test statistic $T$ has to be adapted to account for the different record lengths of data. First let $\mathbf{1}_{i}=(1_{2,i},\ldots,1_{d,i})$ be a $(d-1)$-dimensional vector with $1_{j,i}=0~(1_{j,i}=1)$ if $Z^{N}_{j,i}$ is missing (observed) respectively. Consider a particular vector $\mathbf{Z}_i^{N}$ with missing vector $\mathbf{1}_{i}$ 
where $d_{i}$ elements of $\mathbf{Z}^{N}_{i}$ are observed, i.e., $d_{i}=\mbox{sum}(\mathbf{1}_{i})$ with $0\leq d_{i}\leq d-1$, then
$\mathbf{Z}_i^{N}\sim\mbox{MVN}(0,\Sigma_i)$, where $\Sigma_{i}=\mathbf{1}_{i}\Sigma\mathbf{1}^{'}_{i}$ with $\mbox{dim}(\Sigma_i)=d_{i}\times d_{i}$. By defining
\vspace{-0.8cm}
\begin{center}
\begin{equation}
T_i=\mathbf{Z}_i^{N} \Sigma_i^{-1} (\mathbf{Z}_i^{N})^{'},\nonumber
\end{equation}
\end{center}
it follows that $T_{i}$ has a $\chi_{d_i}^2$ distribution with $\mbox{E}[\chi_{d_i}^2]=d_i$ and $\mbox{Var}[\chi_{d_i}^2]=2d_i$. 
We can define the adapted test statistic of Gaussianity to be
\vspace{-1cm}
\begin{center}
\begin{equation}
T^{*}=\frac{1}{\sqrt{n_{v}}}\sum_{i=1}^{n_{v}}\frac{T_i-d_i}{\sqrt{2d_i}}
\label{eq:Tmis}
\end{equation}
\end{center}
where $n_{v}$ corresponds to the number of observations of $\mathbf{Z}^N$. If a particularly large value of $T^{*}$ is observed then there is a deviation away from the assumption of multivariate normality. The sampling distribution of $T^{*}$ under the null hypothesis for a given pattern of missing data is easily derived by Monte Carlo methods, but has been constructed to have $\mathbb{E}(T^{*})=0$ and $\mathbb{V}\mbox{ar}(T^{*})=1$ under the null hypothesis of the Gaussian copula whatever the missingness pattern, provided $\min(d_{1},\ldots,d_{n_{v}})\geq 1$ and $\Sigma$ is known.

\subsection{Handling missing values}\label{sec:MisCEVM}
The methods given in \citet{HefTawn2004} only consider vectors of complete observations so with any missing data the method will be highly inefficient. The data-usage efficiency can be defined as $100\sum^{n}_{i=1}\mathbbm{1}\left(d_{i}=d-1\right)/n$ with 
$\mathbbm{1}$ being the indicator function and 
$d_{i}$ as defined in Section \ref{sec:TeGa}. \citet{Keef2009} developed a strategy to replace each missing variable by a sample of $m$ replicates generated from a $d-1-d_i$ dimensional Gaussian approximation for the conditional distribution of the missing $\mathbf{Z}^{N}_{i}$ given the observed $\mathbf{Z}^{N}_{i}$ elements for all $i$ with $d_{i}<d-1$. This approach has major computational problems when more than a small number of missing values are present as it requires $w\sum_{i=1}^{n_v} (d-1-d_i)$ simulations, where $w$ needs to be reasonably large to remove Monte Carlo noise, e.g., $w\in(100,1000)$. This approach is subsequently referred to as the infill approach.

We propose using our Gaussian copula model to give a statistically and computationally efficient approach. 
Equation~\eqref{eq:NorRes} is used to transform the $\mathbf{Z}$ variables, on their original margins, 
to $\mathbf{Z}^N$ on Gaussian margins. Concurrent pairs of observations of $\mathbf{Z}^N$ are used to estimate the correlation parameters provided that datum exists for a given $Z_i^N$ and $Z_j^N$ pair. This gives the following estimated correlation matrix $\hat{\Sigma}$, with $(i,j)$th entry of $\hat{\rho}_{i,j}$ being
\vspace{-0.6cm}
\begin{center}
\begin{equation}
\hat{\rho}_{ij}=\frac{\sum_{k=1}^{n_{v}}1_{i,k}1_{j,k}(z_{i,k}-\bar{z_{i}})(z_{j,k}-\bar{z_{j}})}{\sqrt{\sum_{k=1}^{n_{v}}1_{i,k}1_{j,k}(z_{i,k}-\bar{z_{i}})^{2}\sum_{k=1}^{n_{v}}1_{i,k}1_{j,k}(z_{j,k}-\bar{z_{j}})^{2}}}\nonumber,
\end{equation}
\end{center}
with $\bar{z}_{i}=\sum_{k=1}^{n_{v}}1_{i,k}z_{i,k}/\sum_{k=1}^{n_{v}}1_{i,k}$ and similarly for $\bar{z}_{j}$. When there are no concurrent data for the pair $(i,j)$, i.e., $\sum_{k=1}^{n_{v}}1_{i,k}1_{j,k}=0$, then a covariate model or prior information can be used to give an estimate. As the correlation matrix is estimated for non-overlapping data sets, there is a possibility that the resulting estimated correlation matrix $\Sigma$ is not positive semi--definite. However, there are eigen-decomposition methods that can solve this problem by giving the nearest positive-definite matrix $\tilde{\Sigma}$ to $\hat{\Sigma}$ that maintains unit diagonals \citep{Franklin2012}. 

\subsection{Connections with other models}\label{sec:connect}

There have been some Gaussian assumptions made in other work using the \citet{HefTawn2004} model, but that differs from what is proposed here. In the original \citet{HefTawn2004} paper for the inference of the regression parameters ($\boldmath{\alpha}, \boldmath{\beta}$) a pseudo likelihood is constructed with independent Gaussian residuals, but for subsequent inference on $\mathbf{Z}$ this assumption was then dropped. So there is in fact no overlap with the approach in \citet{HefTawn2004}. Motivated by early findings in this paper, in a spatial setting \citet{Tawn2018} assume that $\mathbf{Z}$ is a realisation from a Gaussian process at a set of sites, so there they make an assumption of marginal Gaussianity for $\mathbf{Z}$ in addition to the Gaussian copula we assume. In that paper there is no discussion on how to assess the Gaussian copula model or why it may be appropriate. This is what this paper does.

There is a question of whether our model is reasonable at all.  In fact $\mathbf{Z}$ is multivariate Gaussian for two very widely used copula. Specifically, it arises for the asymptotic dependent multivariate extreme value H{\"u}sler-Reiss distribution copula \citep{Husler1989} with $(\alpha_i,\beta_i)=(1,0)$ for all $i=2, \ldots ,d$, see \citet{Engelke2015}, and for the the asymptotically independent Gaussian copula with $(\alpha_i,\beta_i)=(\rho_{1i}^{2},1/2)$ for all $i=2, \ldots ,d$, see \citet{HefTawn2004}.

\section{Simulation Algorithm and Joint Event Estimation}\label{sec:SimAg}
\subsection{Simulation of extreme events}\label{sec:simEve}
The procedure to simulate from our model for $\mathbf{R}$, assuming that its first component is large, 
is an adaptation of the algorithm in \citet{HefTawn2004} and \citet{Jonathan2013}. Firstly we define 
$q_{i,p}$ as the $p$th quantile of $R_i$, thus $F_i(q_{i,p})=p$. The aim is then to simulate $\mathbf{R}~|~R_1>q_{1,p}$.
On Laplace margins this corresponds to simulating $\mathbf{Y}~|~Y_1>v_{p}$, where $v_p=\log[2(1-p)]$. Here we assume $p$ is sufficiently large so that $v_p>v$, where $v$ is the dependence threshold described in Section \ref{sec:ConDep}.

The steps of the simulation procedure are outlined as follows:
\begin{enumerate}
\item Simulate $\mathbf{Z}^{N}$ from a standard $(d-1)$-dimensional Gaussian distribution with correlation matrix $\hat{\Sigma}$ (as defined in \eqref{eq:NorRes}).
\item Transform $\mathbf{Z}^{N}$ marginally through a 1-dimensional kernel smoothed distribution functions to produce a sample of residuals $\mathbf{Z}=(Z_2, \ldots, Z_d)$, i.e., $Z_i\sim{\hat{G}}^{-1}_i(\Phi(Z_i^N))$ for $i=2,\ldots ,d$.
\item Independent of $\mathbf{Z}^{N}$ draw a value of the conditioning variable $Y_{1}$ from a standard Exponential distribution above $v_{p}$, e.g.,  $Y_{1}=v_{p}+Y^{*}_{1}$, where $Y^{*}_{1}\sim\mbox{Exp}(1)$.
\item Derive the simulated value of the conditioned variates $\mathbf{Y}_{-1}$, which is a function of $Y_{1},~\mathbf{Z}$ and the estimated dependence parameters $(\hat{\bm{\alpha}},~\hat{\bm{\beta}})$, via  
\begin{equation}
\mathbf{Y}_{-1}=\hat{\bm{\alpha}}Y_{1}+Y^{\hat{\bm{\beta}}}_{1}\mathbf{Z},~\mbox{for }Y_{1}>v_{p}.\nonumber
\end{equation}
This gives a sample of $\mathbf{Y}=(Y_{1},\mathbf{Y}_{-1})$ with $Y_{1}>v_{p}$.
\item The inverse of the probability integral transform, as given in equation \eqref{eq:LapTr}, can be used to transform $\mathbf{Y}$ back to its original margins of $\mathbf{R}=(R_{1},\ldots,R_{d})$, 
with $R_1>q_{1,p}$.
\end{enumerate}

In the simulation of spatially consistent extreme events, we want to ensure that events are simulated conditional on $\mathbf{R}$ being extreme for at least one location. We adopt the model of \citet{Keef2013b} that generates an extreme event conditional on the event $\left\{\max(F_{1}(R_{1}),\ldots,F_{d}(R_{d}))>p\right\}$ with $p$ near $1$, or equivalently $\left\{ \exists i=1,\ldots,d:R_{i}>q_{i,p} \right\}$. After transformation to Laplace margins this corresponds to simulating $\max\left(Y_{1},\ldots,Y_{d}\right)>v_{p}$. To be able to simulate from this conditional distribution using the previous algorithm for simulating from $\mathbf{Y}|Y_{1}>v_{p}$, we need to determine the conditioning gauge for each event. The approach is to first simulate
$I^{p}=\argmax\left\{\mathbf{Y}~|\max\left\{Y_{1},\ldots,Y_{d}\right\}>v_{p}\right\}$, with 
\begin{eqnarray*}
\mathbb{P}(I^p=j)&=&\frac{\mathbb{P}(Y_j=\max(Y_1, \ldots ,Y_d), Y_j>v_p)}{\sum_{k=1}^d\mathbb{P}(Y_k=\max(Y_1, \ldots ,Y_d), Y_k>v_p)}\\
&=&\frac{\mathbb{P}(Y_j=\max(Y_1, \ldots ,Y_d)\mid Y_j>v_p)}{\sum_{k=1}^d\mathbb{P}(Y_k=\max(Y_1, \ldots ,Y_d)\mid Y_k>v_p)},
\end{eqnarray*}
where here each of these conditional probabilities can be estimated from our models for $\mathbf{Y}|Y_k>v$, for $k=1,\ldots ,d$. Finally if $I^p=j$ then apply the above algorithm for $\mathbf{Y}|Y_{1}$ with the index $1$ replaced by $j$ and this point is rejected if $\max\left(\mathbf{Y}_{-j}\right)>Y_j$, i.e. steps 1-5 need repeating until for the selected gauge, $j$, we have $\max\left(\mathbf{Y}_{-j}\right)<Y_j$.

\subsection{Estimation of joint extreme events}\label{sec:jtext}
In many applications, such as the design of flood defence schemes or assessing potential flood losses over an insurance portfolio, interest lies in accurately estimating the probability of rare events across a number of spatial locations or environmental hazards. The Monte Carlo methods described in Section~\ref{sec:simEve} are the most effective way to estimate many extreme events, however as was noted in Section~\ref{sec:Int} there are major limitations with these methods for events which are rare relative to the marginal probability for the conditioning variable. Estimation of these probabilities require a more careful analysis, which we can achieve for the first time here due to our semi-parametric residual distribution model choice. We will illustrate the estimation for both these types of events. 

Firstly consider an event $A$ which is extreme in the sense that at least $R_1$ is extreme. Then there exists a value of $p$, near $1$ such that $A\subset [q_{1,p},\infty)\times (\infty, \infty)^{d-1}$. It follows that 
\begin{eqnarray*}
\mathbb{P}(\mathbf{R} \in A) & = & \mathbb{P}(R_1>q_{1,p}) \mathbb{P}(\mathbf{R} \in A~|~R_1>q_{1,p}) \\
& = & (1-p)\mathbb{P}(\mathbf{R} \in A~|~R_1>q_{1,p}).
\end{eqnarray*}
An estimate of this joint probability is given by
\[
\hat{\mathbb{P}}(\mathbf{R} \in A)  = (1-p)\sum_{t=1}^{\ell} \mathbbm{1}(\tilde{\mathbf{R}}_t \in A)/\ell
\]
where $\tilde{\mathbf{R}}_{1},\ldots,\tilde{\mathbf{R}}_{\ell}$ are independent and identically distributed values simulated from $\mathbf{R}|R_1>q_{1,p}$ and $\ell$ is the number of the simulations. However if $\{R_i; i\in C\}$, with $1 \in C$, is asymptotically independent then as $\chi_{C}=0$ the conditional probability that is being estimated by the Monte Carlo methods above is near zero if $A\subset  \prod_{i\in C}(q_{i,p},\infty)$. For sets such as $A$ it is better to exploit the Gaussian copula structure and express the result through an integral for which standard numerical integration methods can be used. Specifically for $A=  \prod_{i\in D} (q_{i,p_i},\infty)$, with $p_1$ near 1, the model gives
\vspace{-0.7cm}
\begin{center}
\begin{eqnarray}
\mathbb{P}\left(R_{1}> q_{1,p_1},\ldots,R_{d} > q_{d,p_d}\right) & = & \mathbb{P}\left(Y_{1}> y_1,\ldots,Y_{d} > y_d\right)\nonumber\\
&=&\int^{\infty}_{y_1}\mathbb{P}\left(\mathbf{Y}_{-1}> \mathbf{y}_{-1} | Y_{1}=s\right)f_{Y_1}(s)ds\nonumber\\
&=&\int^{\infty}_{y_1}\mathbb{P}\left(\boldsymbol{\hat{\alpha}}Y_{1}+Y^{\boldsymbol{\hat{\beta}}}
_{1}\mathbf{Z}> \mathbf{y}_{-1} | Y_{1}=s\right)\frac{1}{2}\exp(-s)ds\nonumber\\
&=&\int^{\infty}_{y_1}\mathbb{P}\left(\mathbf{Z}> \frac{\mathbf{y}_{-1}-\boldsymbol{\hat{\alpha}}s}{s^{\boldsymbol{\hat{\beta}}}} \Huge{|} Y_{1}=s\right)\frac{1}{2}\exp(-s)ds\nonumber\\
&=&\int^{\infty}_{y_1}\mathbb{P}\left(\mathbf{Z}^{N}> \Phi^{-1}\left(\tilde{\mathbf{G}}\left(\frac{\mathbf{y}_{-1}-\boldsymbol{\hat{\alpha}}s}{s^{\boldsymbol{\hat{\beta}}}} \right)\right)\Huge{|} Y_{1}=s\right)\frac{1}{2}\exp(-s)ds\nonumber\\
&=&\int^{\infty}_{y_1}\bar{\Phi}_{d-1}\left(\Phi^{-1}\left(\tilde{\mathbf{G}}\left(\frac{\mathbf{y}_{-1}-\boldsymbol{\hat{\alpha}}s}{s^{\boldsymbol{\hat{\beta}}}} \right)\right),\tilde{\Sigma}\right)\frac{1}{2}\exp(-s)ds
\label{eq:jpest}
\end{eqnarray}
\end{center}
where $\tilde{\mathbf{G}}(\mathbf{z})=(\tilde{G}_{2}(z_{2}),\ldots,\tilde{G}_{d}(z_{d}))$, $\mathbf{y}_{-1}=(y_2, \ldots,y_d)$ with $y_i$ the $p_i$th quantile of a Laplace distribution, and $\bar{\Phi}_{d-1}\left(.,\Sigma\right)$ is the joint survivor function of the standard multivariate Gaussian variable with correlation matrix $\Sigma$. This result allows us to reduce the complexity of the $(d-1)$-dimensional integral calculation of rare event probabilities through the direct evaluation of the multivariate Gaussian joint survivor function and a 1-dimensional integral.
\section{Simulation Study}\label{sec:SimSt}
To assess the performance of our proposed Gaussian copula approach, for modelling the joint distribution of the residuals in the conditional multivariate extremes model, we undertake a simulation study to compare it with the empirical approach of \citet{HefTawn2004} and with an approach using a multivariate kernel density estimate
\vspace{-1cm}
\begin{center}
\begin{equation}
\hat{G}(\mathbf{z})=\frac{1}{n_{v}}\sum^{n_{v}}_{i=1}\Phi_{d-1}\left(\mathbf{z}|\mathbf{z}_{i},\mathbf{H}\right),
\label{eq:jointkern}
\end{equation}
\end{center}
where the $i$th kernel is Gaussian with mean $\mathbf{z}_{i}$ and $\mathbf{H}$ is a positive definite bandwidth matrix \citep{Wand1994} and $\{\mathbf{z}_{1}, \ldots , \mathbf{z}_{n_v}\}$ are the observed residuals. The methods are compared via their estimation of the probability
\vspace{-1cm}
\begin{center}
\begin{equation}
\gamma_{d}=\mathbb{P}\left(R_{1}> q_{1,p},\ldots,R_{d} > q_{d,p}\right)
\label{eq:jpevent}
\end{equation}
\end{center}
with $p=0.99,0.998$ and $0.999$. 

Data are simulated from a symmetric multivariate extreme value logistic distribution \citep{Tawn1990}, with dependence parameter $\delta\in(0,1]$ with the lower and upper limits for $\delta$ corresponding in perfect dependence and independence respectively. For the symmetric logistic distribution and a given dimension $d$, the true probability of equation~\eqref{eq:jpevent} is $\gamma_{d}=\sum^{d}_{m=0}{d \choose m}\left(-1\right)^{m}p^{m^{\delta}}$. For all $\delta<1$ the variables are asymptotically dependent, i.e., $\chi_D>0$, and hence 
parameters of the \citet{HefTawn2004} model are $\boldmath{\alpha}=\boldmath{1}$ and $\boldmath{\beta}=\boldmath{0}$. Furthermore, for this distribution the true copula for $\mathbf{Z}$ is not Gaussian, so our model gives a mis-specification. We consider  $d=5, 10$ and $20$ with $\delta=0.75$ (results with $\delta=0.5$ are not reported but are similar) and a sample size of 5000 with 25 replicated data sets and a 0.98 dependence threshold corresponding to 100 observations being in the joint tail region. Correctly in each case, we find that there is strong evidence to reject the Gaussian copula assumption, at a 5\% level, when using the test statistic~\eqref{eq:Tmis} for each of our simulations. Despite this we proceed to using the Gaussian copula model to see if this mis-specification is important for inference. 

Table~\ref{tab:delt075} shows results for $d=5$ where the regression parameters are both set to their true values and when they are estimated. For this relatively low dimensional case all three methods perform broadly similarly both in terms of their point estimates and bootstrap based $95\%$ confidence intervals, with all intervals containing the truth. Despite its clear mis-specification, the Gaussian copula method gives estimates that are closest to the truth in all 6 cases. Also we see that the multivariate kernel approach performs worst (underestimating) in all cases. 

Furthermore, note that getting good knowledge of the regression parameters $(\boldmath{\alpha}, \boldmath{\beta})$ is more important that the choice of distributional model for $\mathbf{Z}$. This feature is interesting given that much of multivariate extreme value inference has focussed on assuming asymptotic dependence (fixing the regression parameters) and effectively only estimating $\mathbf{Z}$ in different ways. These results suggest that focus of attention has been mis-placed.
\begin{center}
\begin{table}[!h]
\centering
\begin{tabular}{|p{3.8cm}|p{2.5cm}|p{2.5cm}|p{2.5cm}|}
\hline
\textbf{Marginal probability}&\textbf{0.99}&\textbf{0.998}&\textbf{0.999}												\\\hline
True joint probability							&1.80						&0.36						&0.18												\\\hline
\multicolumn{4}{|c|}{\textit{True regression dependence parameters}}														\\\hline
Heffernan and Tawn	&1.97 (1.50,2.48)	&0.39 (0.30,0.50)	&0.20 (0.15,0.25)												\\\hline
Multivariate kernel	&1.63 (1.22,2.03)	&0.32 (0.24,0.41)	&0.16 (0.12,0.20)				\\\hline
Gaussian copula			&1.90 (1.44,2.30)	&0.38 (0.28,0.46)	&0.19 (0.14,0.23)				\\\hline
\multicolumn{4}{|c|}{\textit{Estimated regression dependence parameters}}									\\\hline
Heffernan and Tawn	&1.38 (1.08,1.87)	&0.18 (0.05,0.26)	&0.07 (0.01,0.12)				\\\hline
Multivariate kernel	&1.10 (0.85,1.45)	&0.13 (0.04,0.22)	&0.06 (0.01,0.10)				\\\hline
Gaussian copula			&1.46 (1.03,2.00)	&0.20 (0.07,0.31)	&0.09 (0.02,0.14)				\\\hline
\end{tabular}
\caption{The estimates (with 95$\%$ confidence intervals in parenthesis) for the joint event probability 1000$\gamma_{d}$, given in equation \eqref{eq:jpevent}, for $d=5$ with $\delta=0.75$ for a sample of size 5000 from the symmetric logistic distribution.}\label{tab:delt075}
\end{table}
\end{center}

Higher dimensional studies, $d=10$ and $20$, are compared in Table~\ref{tab:highd} with the true regression dependence parameters treated as known to enable easier comparison of the different methods for handling the residuals. The multivariate kernel approach is now clearly failing when $d=10$ and becomes increasingly computationally expensive as $d$ increases and so is omitted from the $d=20$ study.  The empirical approach of \citet{HefTawn2004} and the Gaussian copula approach perform broadly similarly as well, though again where they differ the Gaussian method works best. 
So even in this case with clear mis-specification the proposed Gaussian copula is at least very highly competitive relative to the existing method. It should be noted that when there is either no mis-specification or there are missing data, 
the Gaussian copula approach substantially out-performs the empirical approach of \citet{HefTawn2004}, see
Section~\ref{sec:missingtest} for an example of this.

\begin{center}
\begin{table}[!h]
\centering
\begin{tabular}{|p{3.8cm}|p{2.5cm}|p{2.5cm}|p{2.5cm}|}
\hline
\textbf{Marginal probability}&\textbf{0.99}&\textbf{0.998}&\textbf{0.999}												\\\hline
\multicolumn{4}{|c|}{\textit{d=10}}																															\\\hline
True joint probability							&1.39						&0.28						&0.14												\\\hline
\multicolumn{4}{|c|}{\textit{True regression dependence parameters}}														\\\hline
Heffernan and Tawn	&1.49 (0.98,1.87)	&0.30 (0.20,0.37)	&0.15 (0.10,0.19)												\\\hline
Multivariate kernel	&0.79 (0.52,1.05)	&0.16 (0.10,0.21)	&0.08 (0.05,0.11)												\\\hline
Gaussian copula			&1.34 (1.00,1.65)	&0.27 (0.20,0.33)	&0.13 (0.10,0.17)												\\\hline
\multicolumn{4}{|c|}{\textit{d=20}}																															\\\hline
True joint probability							&1.15						&0.23						&0.11												\\\hline
\multicolumn{4}{|c|}{\textit{True regression dependence parameters}}														\\\hline
Heffernan and Tawn	&1.09 (0.84,1.50)	&0.22 (0.17,0.30)	&0.11 (0.10,0.19)												\\\hline
Gaussian copula			&1.12 (0.81,1.33)	&0.22 (0.16,0.27)	&0.11 (0.08,0.13)												\\\hline
\end{tabular}
\caption{The estimates (with 95$\%$ confidence intervals in parenthesis) for the joint event probability 1000$\gamma_{d}$, given in equation \eqref{eq:jpevent}, for $d=10$ and $20$ with $\delta=0.75$ for a sample of size 5000 from the symmetric logistic distribution.}\label{tab:highd}
\end{table}
\end{center}

\section{River Flow Applications}\label{sec:WkEx}
\subsection{Data}
\label{sec:data}
\begin{figure}[!h]
\begin{center}
\includegraphics[scale=0.4]{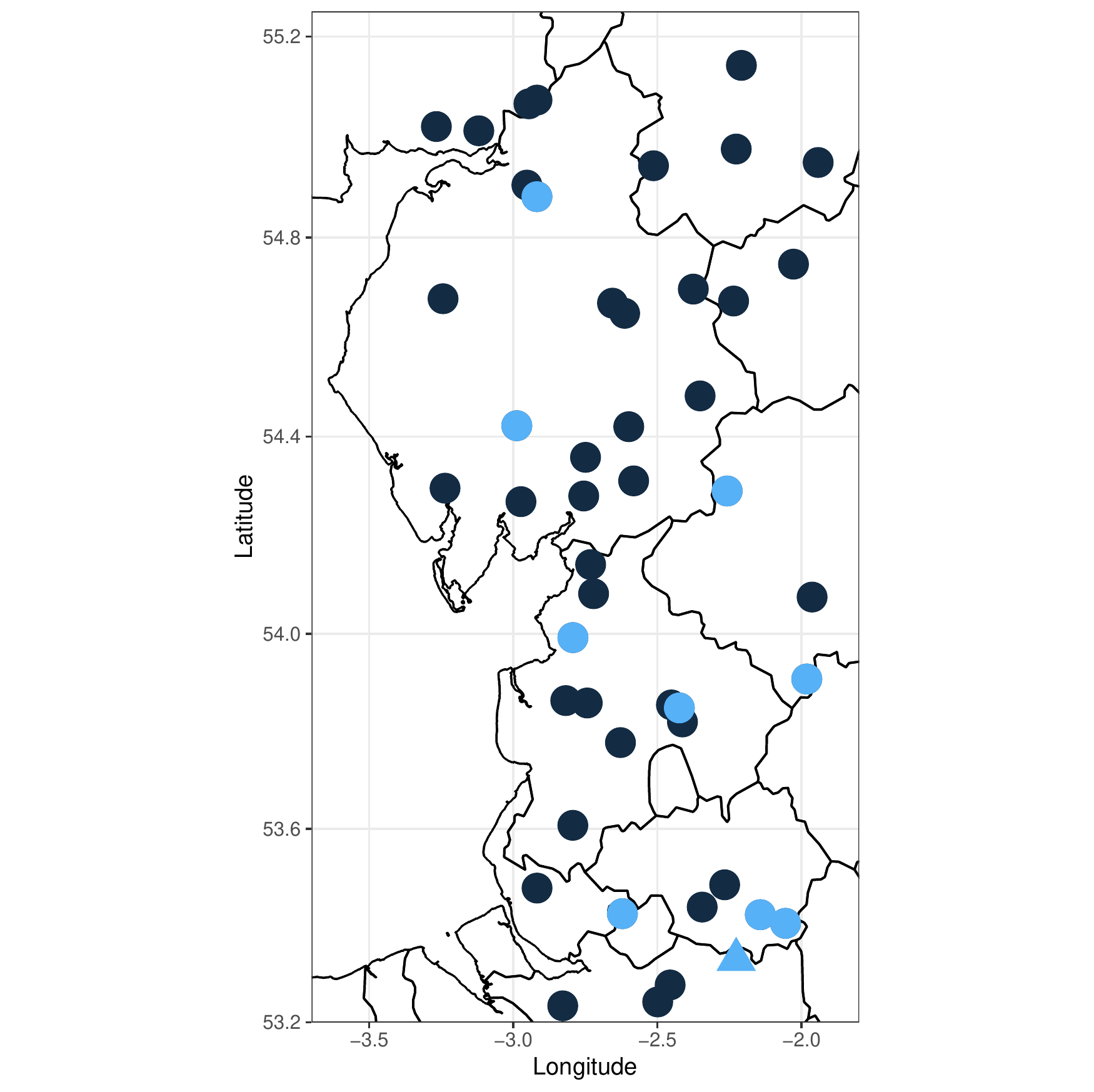}
\caption{Locations of the 46 daily mean river flow gauges situated in the north west of England. The subset of 10 gauges are shown given in light blue. The conditioning station used in estimation of probability $\tau_{m,p}$, defined in equation \eqref{eq:COrdProb}, is represented by a triangle. }
\label{fig:NWNRFA}
\end{center}
\end{figure}

We apply the proposed semi-parametric conditional extreme value model to daily mean measurements of river flow data from the National River Flow Archive (NRFA) to answer questions typically proposed by flood risk managers. Gauges from the north west region of England were selected and the locations of these are given in Figure~\ref{fig:NWNRFA}; on average each gauge has record length of approximately 30 years. This region has one of the better spatial coverages of data in the UK. The proportion of missing values in the data is relatively low.  The region exhibits varying spatial characteristics, for example due to changing soil types and elevation the behaviour is likely to be very different in Cumbria compared to say Manchester (in the north and south of the region respectively). 
The data set was selected as it has been used for previous spatial flood risk assessments \citep{Lamb2010,Tawn2018,Towe2016} and it is a region badly affected by the 2015 floods, discussed in Sections~\ref{sec:Int} and \ref{sec:stDes}.
For the data, we discuss how our proposed methodology can aid in producing better inferences for rare events at much reduced computational cost and with minimal risk of mis-specification error.

In Section \ref{sec:BaCa} we will illustrate all of the steps of the methodology with a basic case study of 10 sites, then undertake to a full application to 46 gauges in Section \ref{sec:RegUK}. We see the 10 site study as important as it lets us look carefully at some of the features of the modelling/inference without getting lost in the volume of the data. In particular, we can look at what happens when large portions of the data are missing.
To help investigate how our methods work in the basic case study we estimate probabilities of extreme events for two data sets. The original data set, denoted F, has $1\%$ missing ($0.5\%$ are missing conditional on the first site being large), with a missingness pattern that allows use of \citet{HefTawn2004} and the infill approach of \citet{Keef2009}. The second data set, denoted M has 28$\%$ removed to missing status in such a way that no complete observations are available ($30\%$ are missing conditional on the first site being large). In both of the analyses the conditioning site is the same and can be identified by the triangle in Figure~\ref{fig:NWNRFA}. For the full application considering 46 gauges, in Section \ref{sec:RegUK}, 2$\%$ of the data are missing.

\subsection{Basic case study}\label{sec:BaCa}

\subsubsection{Assessing the Gaussian copula}
\begin{figure}[!h]
\begin{center}
\includegraphics[scale=0.4]{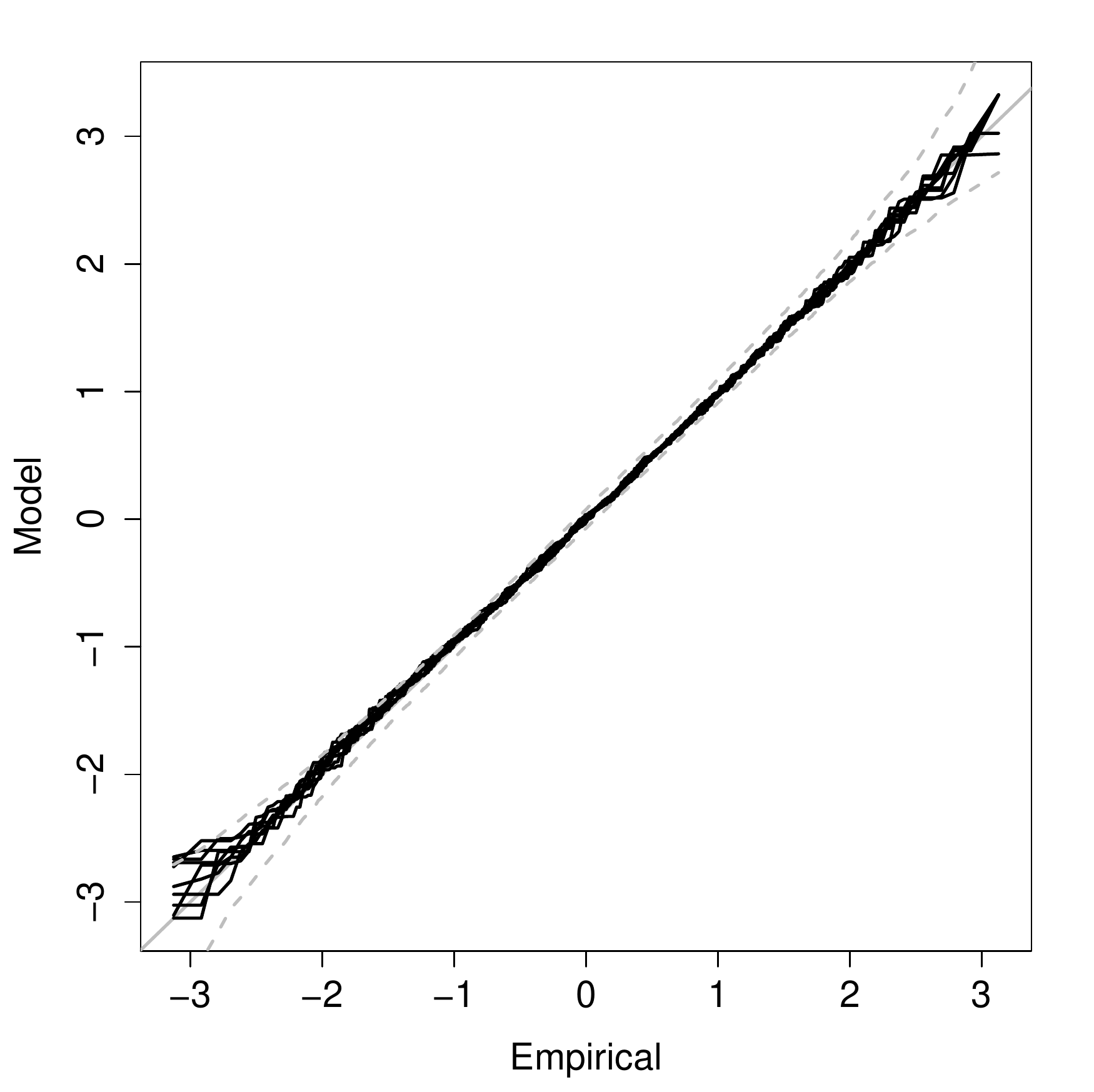}
\caption{Pooled marginal QQ plots of $\mathbf{Z}^{N}=(Z^{N}_{2},\ldots,Z^{N}_{10})$.}
\label{fig:MargKern}
\end{center}
\end{figure}

First we use the original data set to assess our modelling assumptions for these data. Conditioning on $R_{1}$ being large, we focus on studying the behaviour of $\mathbf{Z}^{N}$, the residuals after the marginal transformation to standard Gaussian margins. A check of the assumption of standard Gaussian margins is given in Figure \ref{fig:MargKern}, the empirical quantiles of a standard Normal are plotted against those of the residuals $\mathbf{Z}^{N}$ with this being a pooled QQ plot over all margins and replicates of $\mathbf{Z}^{N}$. The different lines in Figure~\ref{fig:MargKern} for each respective margin of $\mathbf{Z}^{N}$ show that there is no significant deviation away from the line of equality, therefore the marginals satisfy the assumptions for the proposed Gaussian copula model.

\begin{figure}[!h]
\begin{center}
\includegraphics[scale=0.4]{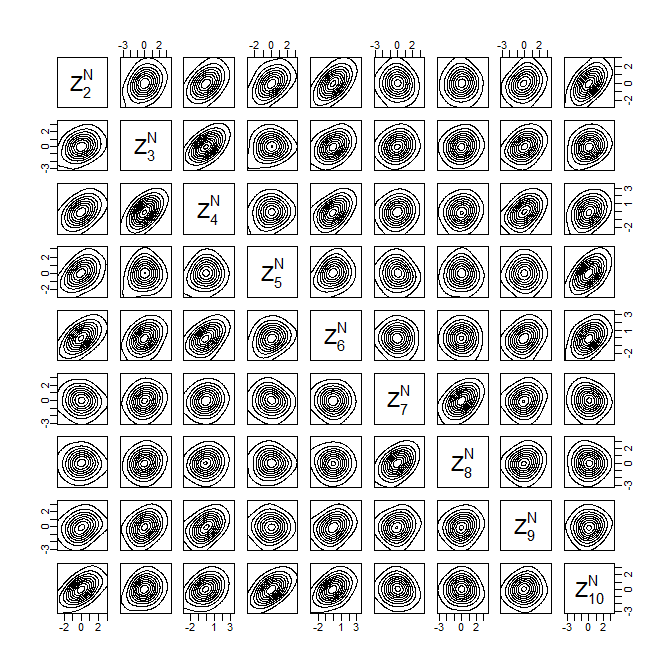}
\caption{Pairwise kernel density estimates for $\mathbf{Z}^{N}=(Z^{N}_{2},\ldots,Z^{N}_{10})$.}
\label{fig:NormRes}
\end{center}
\end{figure}

Pairwise bivariate kernel density estimates for $\mathbf{Z}^{N}$ can be seen in Figure~\ref{fig:NormRes}. From a visual inspection the pairwise dependence seems close to Gaussianity, although in a couple of pairs such as $(Z^{N}_{3},Z^{N}_{5})$ there does seem to be departure away from the expected elliptical contours. Figure~\ref{fig:NormRes} does not help us assess any higher order dependence, and as a result the test for Gaussianity (as given in Section~\ref{sec:TeGa}) is performed to test the assumption of a Gaussian copula more rigorously. The test statistic is calculated using the methodology given in Section~\ref{sec:TeGa}. The \textit{p}-value is calculated to be equal to 0.29, which is greater than the significance level of 0.05. Therefore the assumption of a Gaussian copula seems reasonable. 

\begin{figure}[!h]
\begin{center}
\includegraphics[scale=0.5]{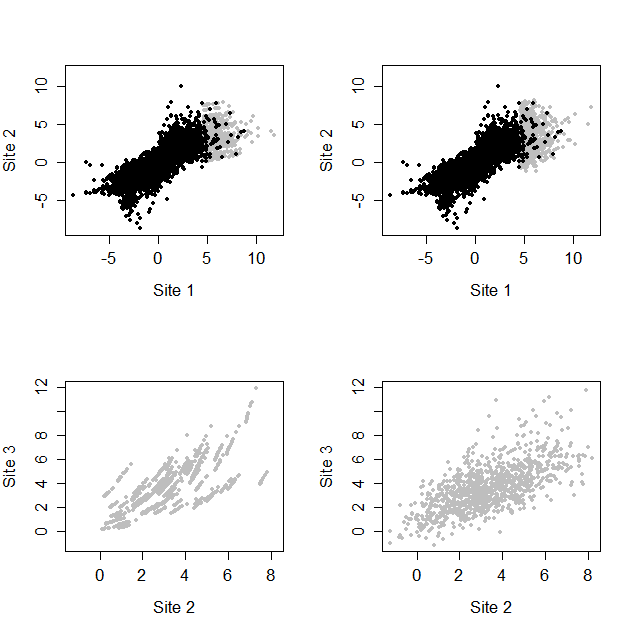}
\caption{Top row: observed (black) and joint behaviour of site 1 and site 2 and simulated (grey) given that an extreme event is observed at site 1. Bottom row: observed (black) and simulated (grey) joint behaviour of site 2 and site 3 given that an extreme event is observed at site 1. Left: the existing method; right: our proposed method. In all figures the data are shown after transformation to standard Laplace margins.}
\label{fig:MisGau}
\end{center}
\end{figure}

Some benefits of the Gaussian copula approach are that the new method is able to interpolate and extrapolate the observed residuals giving simulated events which are not simply deterministic functions of observed events. A comparison of these features of the \citet{HefTawn2004} and Gaussian copula approaches is illustrated in Figure \ref{fig:MisGau}. Under these two approaches Figure \ref{fig:MisGau} (top) shows data and simulations of $Y_{2}|Y_{1}>v_{p}$, (bottom) shows $(Y_{2},Y_{3})|Y_{1}>v_{p}$; both for $p=0.99$. From the top row our proposed approach is seen to give a continuous distribution for $Y_{2}|Y_{1}$ with slightly more variation in $Y_{2}|Y_{1}>v_{p}$. This additional variation, which seems realistic given the extremal behaviour of the observed data set, is due to the use of a kernel smoothed marginal distribution functions for $\mathbf{Z}^{N}$. Similarly, from the bottom row, it can be seen that the simulated joint residuals can differ from observed values, due to the Gaussian copula assumption. Collectively these new features lead to the simulation of a more realistic joint sample with our proposed approach than from the \citet{HefTawn2004} model.
\subsubsection{Conditional probabilities for flood risk management}
\label{sec:missingtest}
In many flood risk management cases, interest lies in determining the spatial extent of any given flood event. One common risk measure that flood managers are interested in is the probability that given a site, site 1 say, exceeds its $p$th quantile that there are then at least $m$ other sites that also exceed their respective $p$th quantile, i.e.,
\vspace{-1cm}
\begin{center}
\begin{equation}
\tau_{m,p}=\mathbb{P}(\#(j=2,\ldots,d:R_{j}> q_{j,p})\geq m~|~R_{1}>q_{1,p})=\mathbb{P}(Y_{(m)}>v_{p}|~Y_{1}>v_{p}),
\label{eq:COrdProb}
\end{equation}
\end{center}
$m=1,\ldots,d-1$ and where $Y_{(m)}$ is the $m$th largest value of $(Y_{2},\ldots,Y_{d})$. Probabilities\\ $\tau_{m,p}~(m=1,\ldots,d-1)$ are useful as they give a clear insight into the spatial extent of a flooding event. If the $p$th quantile is the level of flood defence at all sites, the probability of exactly $m$ other sites being flooded, given site 1 floods is $\tau_{m,p}-\tau_{m+1,p}$.

\begin{center}
\begin{table}[!h]
\centering
\begin{tabular}{|p{2cm}|p{2.5cm}|p{2cm}|p{2.5cm}|p{2cm}|p{2.5cm}|} 
\hline
\textbf{Probability}&\textbf{Heffernan and Tawn (F)}&\textbf{Infill (F)}&\textbf{Gaussian Copula (F)}&\textbf{Infill (M)}&\textbf{Gaussian Copula (M)}\\\hline
$\tau_{5,100}$&4.3 (0.0,12.2)	&4.3 (0.1,11.9)	&4.1 (0.1,12.7)		&4.0 (0.6,15.5)&4.5 (0.3,14.5)			\\\hline
$\tau_{5,500}$&2.9 (0.0,9.6)	&2.9 (0.0,9.7)	&3.0 (0.0,10.4)		&2.4 (0.2,13.7)&3.1 (0.1,13.1)			\\\hline
$\tau_{5,1000}$&2.5 (0.0,8.3)	&2.5 (0.0,8.9)	&2.4 (0.0,9.5)		&2.0 (0.1,12.5)&2.6 (0.1,12.3)			\\\hline
$\tau_{5,10000}$&1.6 (0.0,6.8)&1.6 (0.0,6.8)	&1.6 (0.0,7.8)		&1.0 (0.0,10.9)&1.7 (0.0,10.1)			\\\hline
\end{tabular}
\caption{The estimates (with  95$\%$ confidence intervals in parenthesis) for the conditional probability 100$\tau_{m,T}$, given in equation \eqref{eq:COrdProb}, with $m=5$ using the original (F) and 28$\%$ missing data (M). The $T$ is the probability that corresponds to a specific annual return period. The \citet{HefTawn2004} column corresponds to the conditional extreme value model fitted to all of the data. The modelled infill column refers to the missing values being modelled and infilled into the observed data.}\label{tab:ObsCP}
\end{table}
\end{center}

For $\tau_{m,p}$, given in equation~\eqref{eq:COrdProb} with $m=5$, in Table \ref{tab:ObsCP} we provide a point estimate and associated $95\%$ confidence intervals, obtained by using the parametric bootstrap for a range of return periods. These estimates are compared using the \citet{HefTawn2004} method with two missing value methods 
(infill method of \citet{Keef2009} and our proposed Gaussian copula method). The two data sets denoted F and M are considered, see Section~\ref{sec:data}.

For data set F, all three methods produce very similar estimates. This is not surprising for the \citet{HefTawn2004} and infill methods as for $99\%$ of the data these methods are identical. However for the Gaussian copula we are using the modelled residual copula for all the data that are extreme at the conditioning site, and so to find that the estimate varies so little from that of \citet{HefTawn2004} is particularly pleasing. For the F data, confidence intervals for both the missing data methods are largest due to a combination of the additional Monte Carlo uncertainty and residual marginal distribution smoothing in the respective methods. Here only $1\%$ of the data were missing, so we would not expect to see any clear improvement in using these missing data methods, which use all partially observed components unlike in the \citet{HefTawn2004} method. 

For data set M, it is impossible to obtain estimates from the \citet{HefTawn2004} approach due to there being no observations being made concurrently. What is pleasing to see here is that the two missing data methods give broadly similar estimates to those from data set F. In particular, the Gaussian copula model gives estimates which are very close
to those using the F data sets for all events in Table~\ref{tab:ObsCP} whereas for the infill method the estimates are less self-consistent for the rarer of these events. The confidence intervals of the two methods are approximately the same, which is to be expected as both model the missing values by using a Gaussian copula but handle the computation in different ways. Naturally, the confidence intervals for the M data are larger than the equivalent ones for the F data.  

A critical feature is that the Gaussian copula approach is computationally much quicker even in this basic case. Specifically, the time to get the point estimates using the Gaussian copula is $30\%$ less than the infill method (assuming $\omega=100$), and this efficiency gain improves dramatically as both the number of sites and the proportion of missing data increase.

The probabilities in Table \ref{tab:ObsCP} were estimated through simulation. However, if we were interested in all sites being above a given return level, this corresponds to $m=9$ in equation \eqref{eq:COrdProb}. This probability is incredibly computationally expensive to estimate through Monte Carlo simulation, however the methods developed in Section~\ref{sec:jtext} can provide us with an estimate which avoids Monte Carlo noise, as it obtained using the formulation~\eqref{eq:jpest} divided by $p$, with $d=10$. Table \ref{tab:IntCP} provides estimates of the $\tau_{9,p}$ for the same return periods as in Table~\ref{tab:ObsCP} along with the corresponding numerical integration error.

\begin{center}
\begin{table}[!h]
\centering
\begin{tabular}{|p{2cm}|p{2.2cm}|p{3cm}|} 
\hline
\textbf{Probability}&\textbf{Estimate}			&\textbf{Numerical Error}	\\\hline
$\tau_{9,100}$		&$7.34\times 10^{-5}$		&$6.00\times 10^{-8}$		\\\hline
$\tau_{9,500}$		&$1.19\times 10^{-5}$		&$1.26\times 10^{-8}$		\\\hline
$\tau_{9,1000}$		&$8.58\times 10^{-7}$		&$1.60\times 10^{-8}$		\\\hline	
$\tau_{9,10000}$	&$2.51\times 10^{-12}$	&$1.90\times 10^{-14}$		\\\hline
\end{tabular}
\caption{The estimates (and integration numerical error) for the conditional probability $\tau_{m,p}$, given in expression~\eqref{eq:jpest}, with $m=9$. The table uses the same return periods as in Table~\ref{tab:ObsCP}.}\label{tab:IntCP}
\end{table}
\end{center}

\subsection{Large-scale study}\label{sec:RegUK}
Here the entirety of the north west region of England is considered, this equates to 46 sites in our study. The first modelling step is to fit the conditional extreme value model of \citet{HefTawn2004} conditioning on each of the 46 gauges in turn. For each of these 46 models the estimates of the dependence parameters $\boldsymbol{\alpha}$ and $\boldsymbol{\beta}$ are obtained along with the residuals $\mathbf{Z}$ of the model.

\begin{figure}[!h]
\begin{center}
\includegraphics[width=8cm,height=8cm]{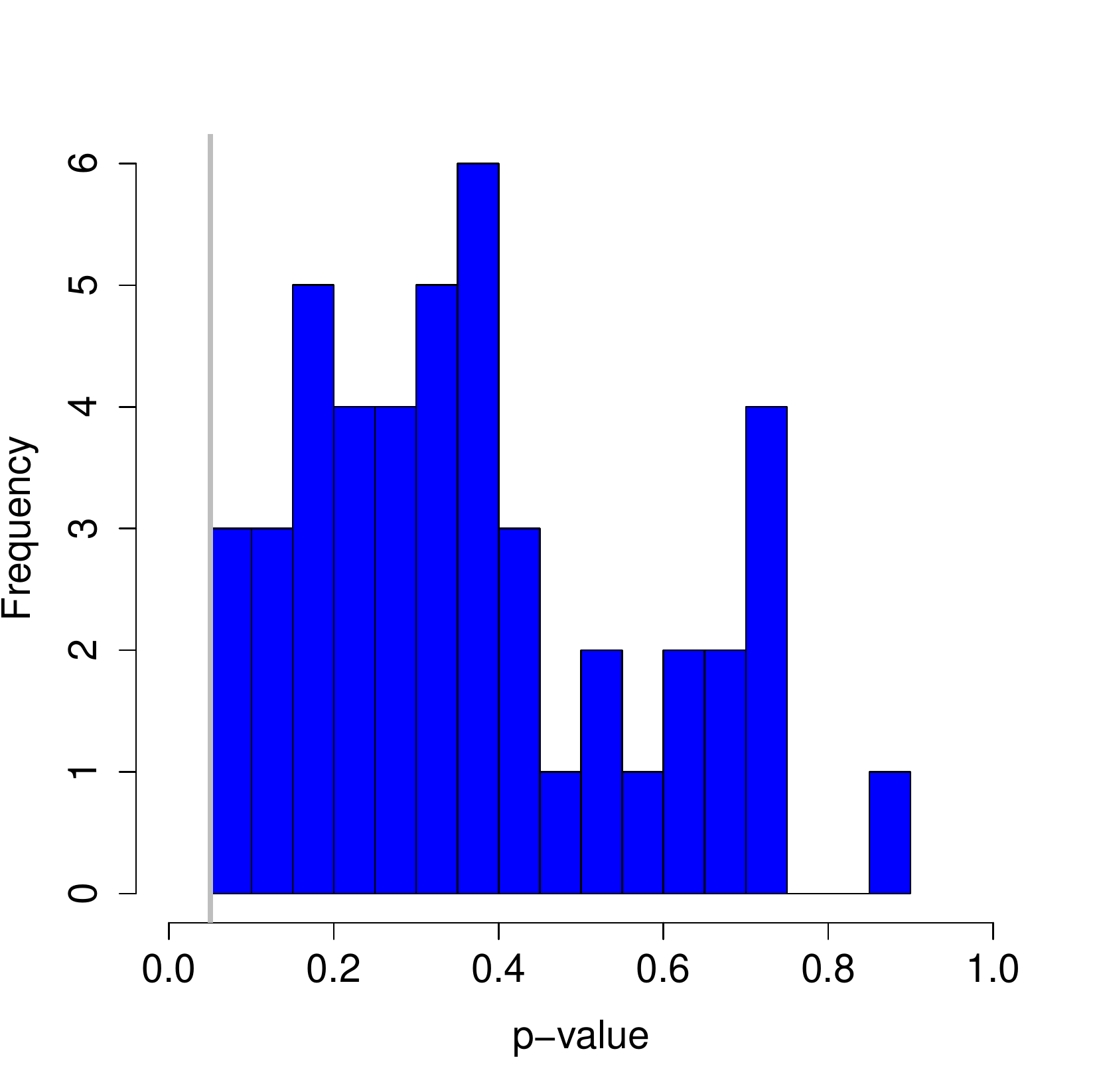}
\caption{Histogram of test statistic $p$ values for the hypothesis of a Gaussian copula for $\mathbf{Z}^{N}$ under 46 different conditioning sites.}
\label{fig:NWTestStat}
\end{center}
\end{figure}

The residuals $\mathbf{Z}^{N}$ of the model are tested to determine whether they can be characterised by using a Gaussian copula. For each conditioning gauge in turn the sampling distribution of the test statistic $T^{*}$, as given in Section \ref{sec:TeGa}, is obtained through Monte Carlo simulation and a $p$-value for a Gaussian copula is derived. Figure~\ref{fig:NWTestStat} shows a histogram of the $p$-values with all of the 46 $p$-values above the 5$\%$ significance level. Therefore, we can conclude that there is no evidence against modelling the residual distribution with a Gaussian copula. Given this conclusion it seems reasonable to use the model-based Gaussian copula for the multivariate residual component of the conditional extreme value model of \citet{HefTawn2004}.

\begin{figure}[!h]
\centering
\begin{subfigure}[c]{.45\linewidth}
\includegraphics[width=\textwidth]{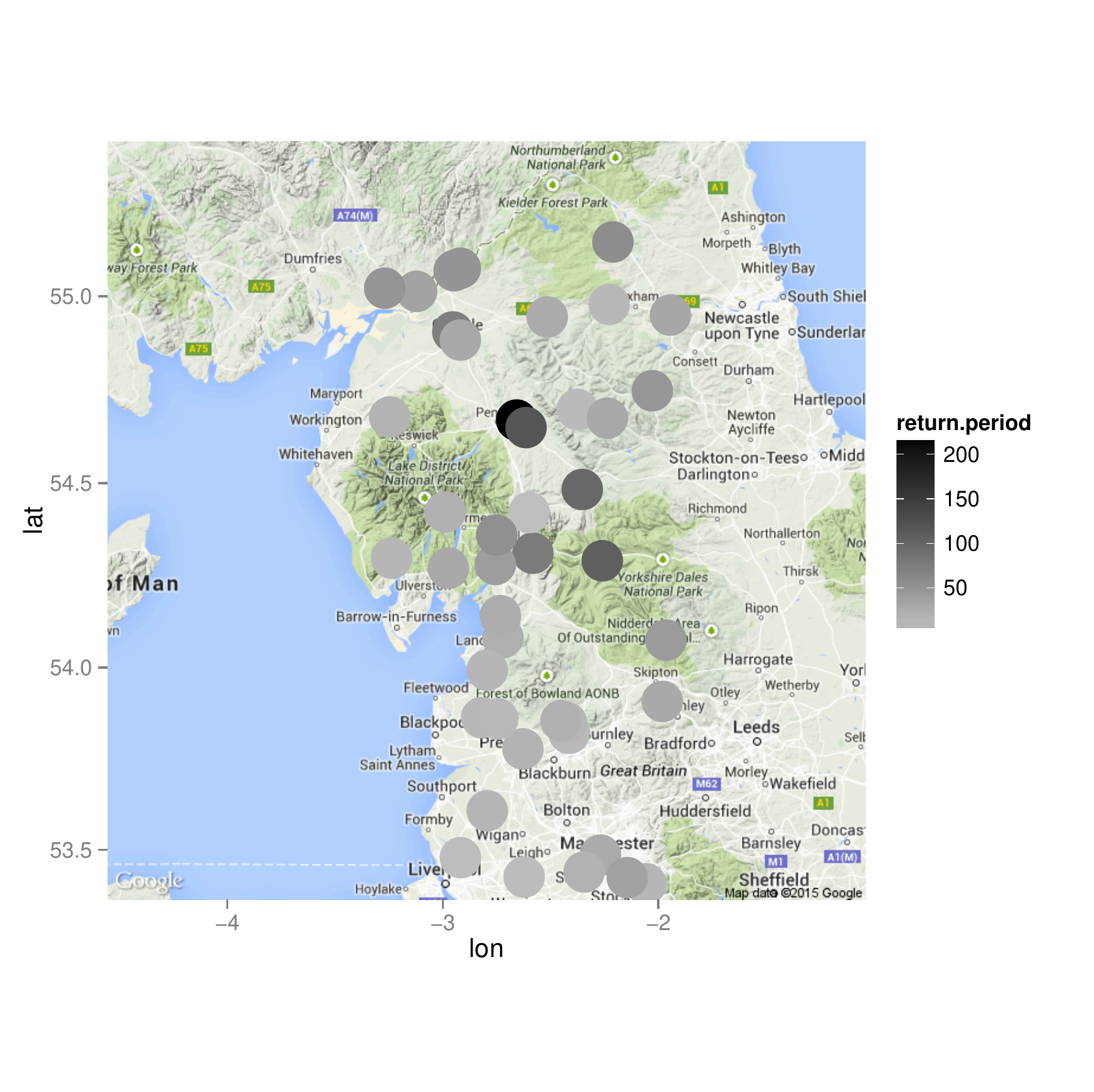}
\label{fig:ev1}
\end{subfigure}
\begin{subfigure}[c]{.45\linewidth}
\includegraphics[width=\textwidth]{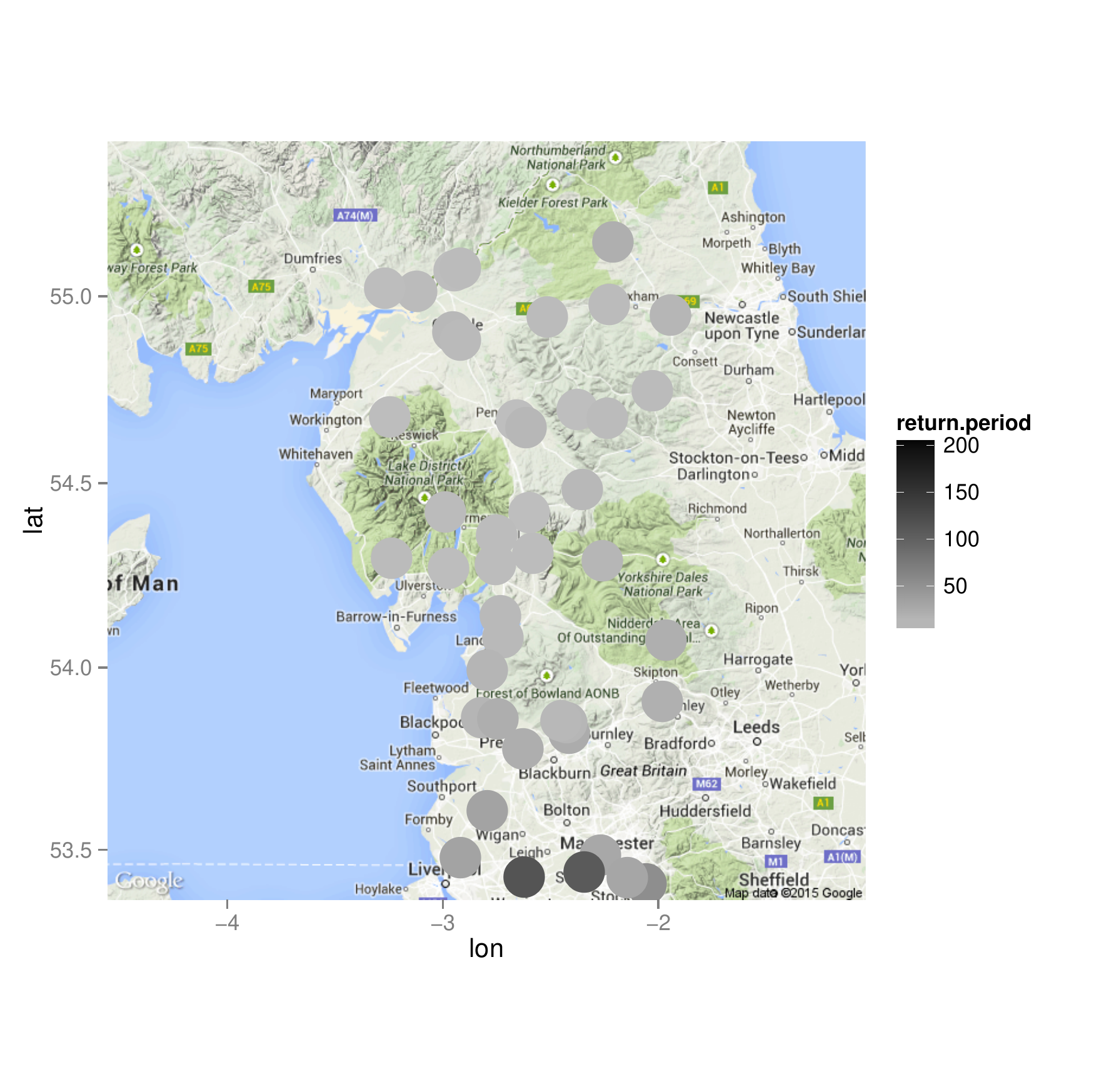}
\label{fig:ev2}
\end{subfigure}
\caption{Two realisations (on the return period scale) from the proposed model, where the conditioning gauge observes at least a 1 in 100 year event, the conditioning sites are: left, in Cumbria; right, close to Manchester.}%
    \label{fig:NFRAEventSet}%
\end{figure}

We can use these models to make extrapolations using the Monte Carlo methods given in Section~\ref{sec:simEve}. These simulations maintain the extremal dependence structure of the observed data set but will also generate events that are larger and more varied than those we have already observed. Two such examples are shown in Figure~\ref{fig:NFRAEventSet} with these illustrating how the spatial structure of an event varies depending on where in the region the event is extreme. The two events have been selected to be extreme at two different sites in the region, in Cumbria and Manchester, in the north and south of the region respectively. In Figure~\ref{fig:NFRAEventSet}(a), when the conditioning location is in Cumbria, there is a much wider spatial impact, than in Figure~\ref{fig:NFRAEventSet}(b), for an event near Manchester. This reflects that when we condition on Cumbria being extreme, relative to Manchester being extreme, the associated $\boldsymbol{\alpha}$ parameters are larger over many more sites, so the spatial extremal dependence is stronger and extreme events in the north of the region are more widespread than those in the south of the region.

\begin{figure}[!h]
\begin{center}
\includegraphics[scale=0.21,keepaspectratio]{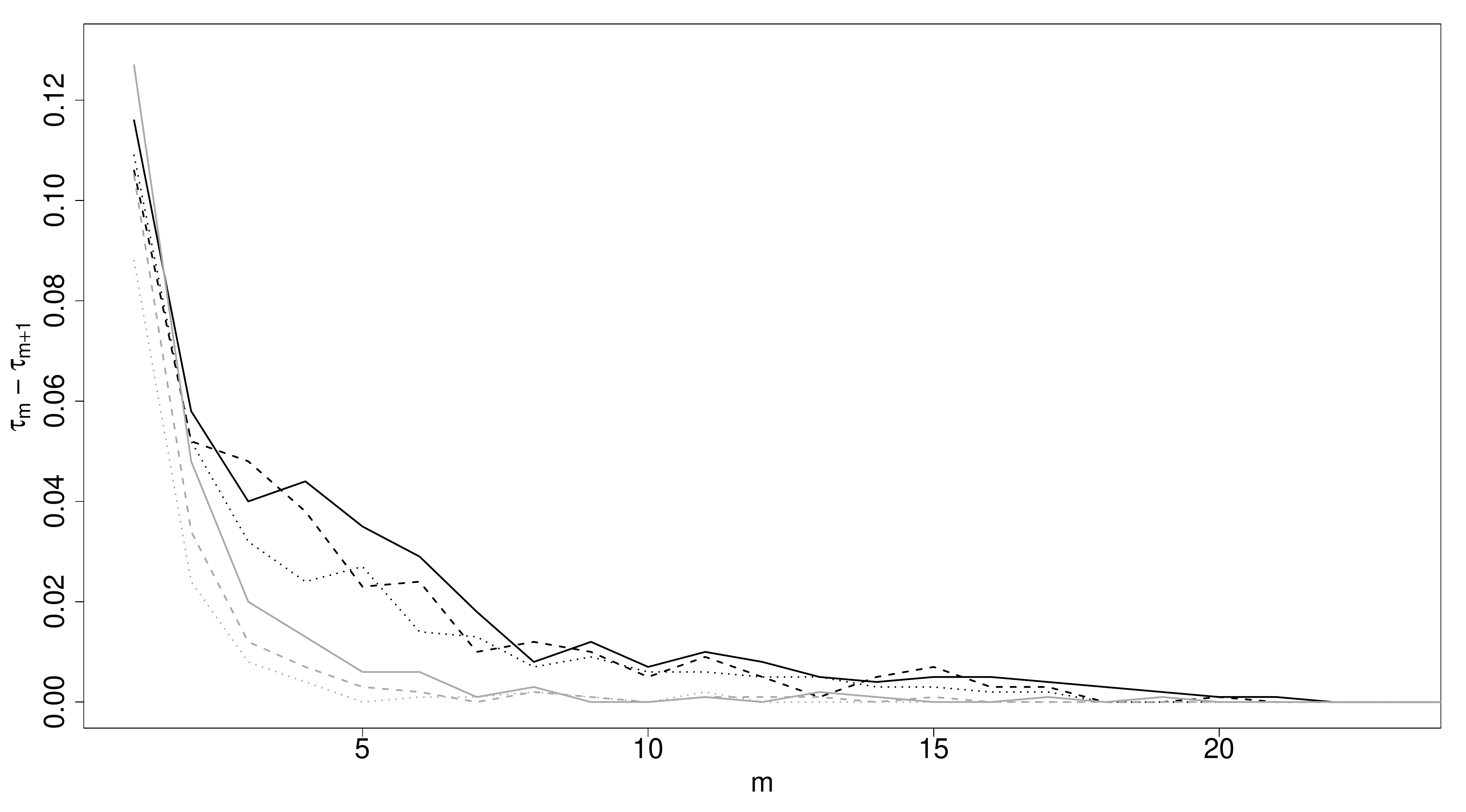}
\caption{Distribution of the number $m$ of other sites that are extreme given the condition site is extreme: the grey lines, conditioning on gauge 69017 near Manchester; black lines conditioning on gauge 74001,  in Cumbria. The solid, dashed and dotted correspond to observing a 100, 1000 and 10000 year event at the respective conditioning site.}
\label{fig:CondProb}
\end{center}
\end{figure}

To further study the varying spatial characteristics of extreme flood events, a conditioning site is selected to have an extreme event and the distribution of the number of other gauges that are also extreme is estimated. This estimated distribution is derived for the same two conditioning sites as in Figure~\ref{fig:NFRAEventSet}. Here the probability of exactly $m$ other gauges is $\tau_{m,p}-\tau_{m+1,p}$, and this is estimated for three return periods. 
Estimates of $\tau_{m,p}-\tau_{m+1,p}$ are compared in Figure \ref{fig:CondProb} for the two conditioning gauges. There is a clear difference in these estimated probabilities. The estimates show that there is greater clustering of flood events when conditioning on the Cumbria site being large. However, some of this clustering could be explained by the fact there are a higher density of gauges in this region. Furthermore, the estimates decay to zero, for $m>1$, at different rates, thus events become more localised as they become more extreme, due to asymptotic independence. 

\subsection{Determining the rarity of the storm Desmond event}\label{sec:stDes}

\begin{figure}[!h]
\centering
\begin{subfigure}[c]{.45\linewidth}
\includegraphics[width=\textwidth]{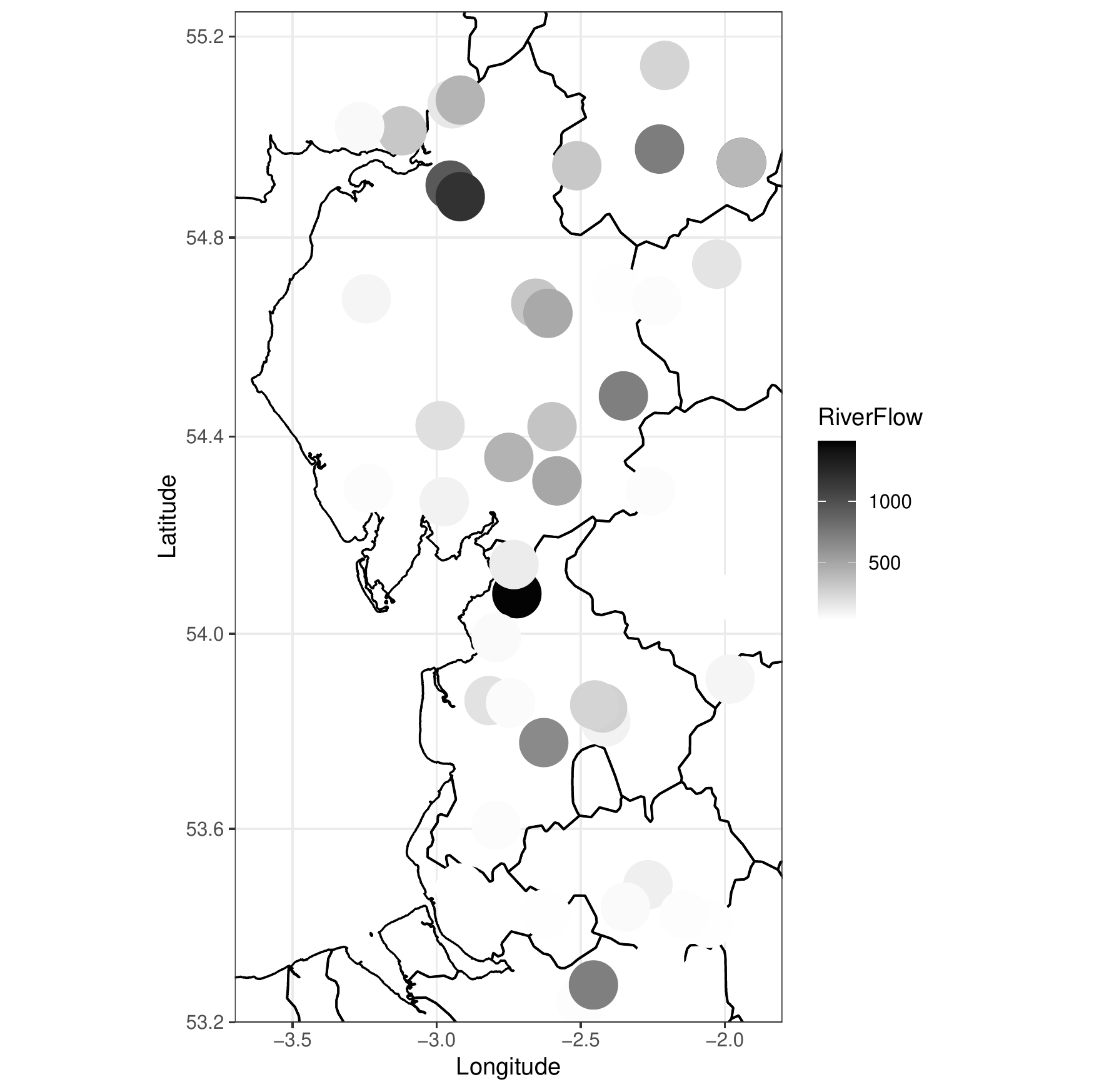}
\label{fig:obs}
\end{subfigure}
\begin{subfigure}[c]{.45\linewidth}
\includegraphics[width=\textwidth]{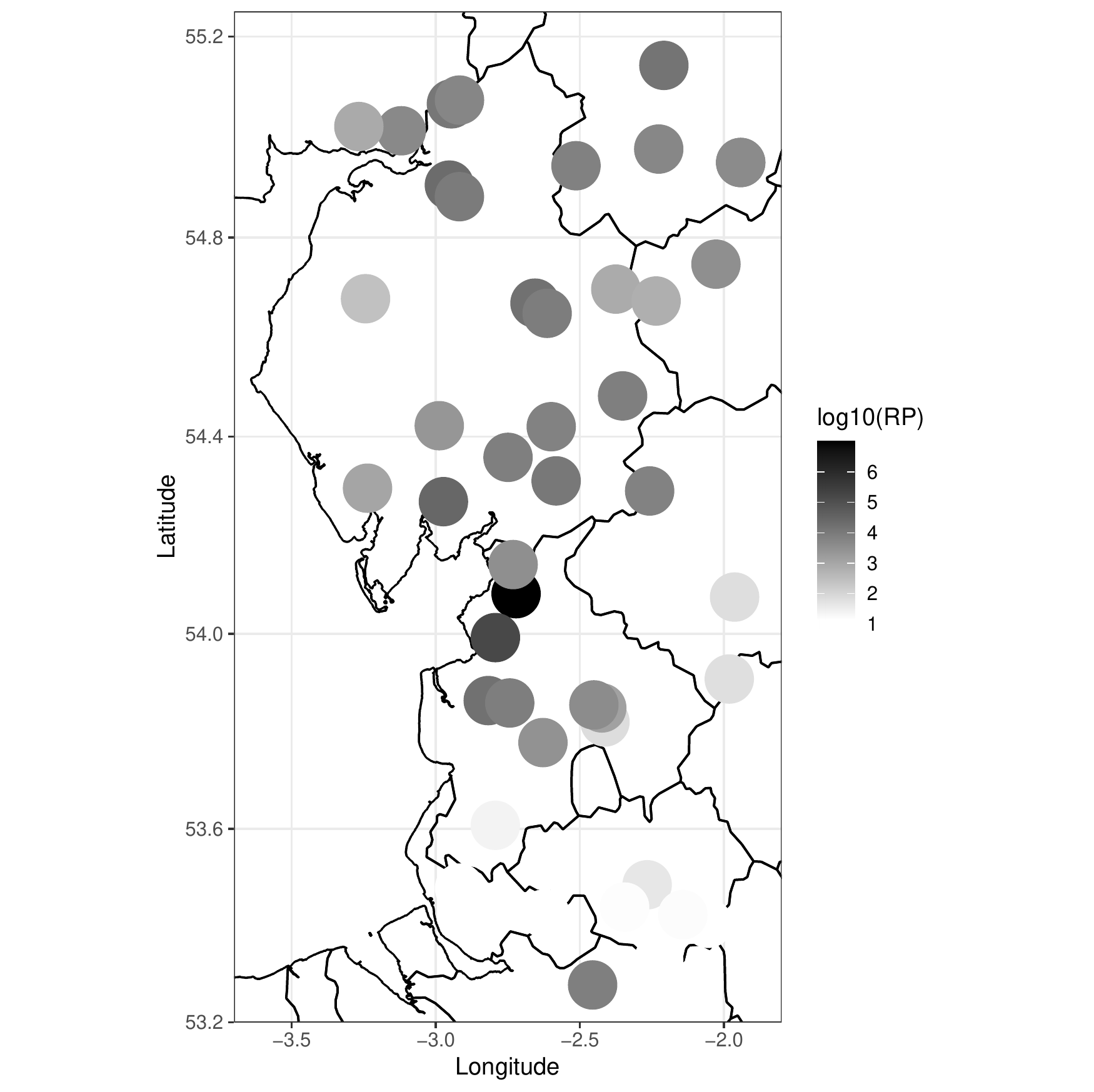}
\label{fig:rp}
\end{subfigure}
\caption{Left: observed daily mean river flows measured in $m^{3}s^{-1}$ from the 5th December 2015 and right: the corresponding marginal return periods for those observed daily mean river flows plotted on the log scale.}%
    \label{fig:Dec2015}%
\end{figure}

The methodology is used to determine the rarity of river flows that were observed on the 5th December 2015 storm Desmond event. This estimate is derived from the daily mean river flow data discussed in Section~\ref{sec:data} with the results presented on a daily scale. The observed daily mean river flows are shown in Figure~\ref{fig:Dec2015}(a) with the largest values observed near Lancaster and Carlisle. However, when we determine the associated estimated marginal return periods, with inference using the GPD tail model~\eqref{eq:CDFDef}, the river flow observed near Lancaster is found to be the most extreme, as shown in Figure~\ref{fig:Dec2015}(b). The marginal observational probability for the Lancaster gauge is estimated to be $3.6\times 10^{-5}$.  Figure~\ref{fig:Dec2015}(b) shows that the event was particularly rare over all Cumbria and northern Lancashire, but it was extreme at only one of the gauges near Manchester in the south of the study region.

In order to determine the probability $\mathbb{P}\left(R_{1}> q_{1,p_1},\ldots,R_{d} > q_{d,p_d}\right)$
of jointly observing river flows over the region which are worse than the 2015 event we use  both the empirical \citet{HefTawn2004} residual approach and our Gaussian copula approach with the joint probability given 
by the integral~\eqref{eq:jpest}. We illustrate the calculations by separately taking the conditioning gauge to the 
Cumbrian gauge, shown in Figure \ref{fig:NFRAEventSet}(a), and the Lancaster gauge, identified by Figure~\ref{fig:Dec2015}(b). 
Using the Cumbrian gauge we estimate the joint probability to be $<1.60\times 10^{-12}$ and $3.70\times 10^{-9}$ 
using the respective methods, whereas these respective estimates become $9.50\times 10^{-10}$ and 
$8.00\times 10^{-9}$ 
using Lancaster. When conditioning on the Cumbrian gauge we can only bound the joint probability using the empirical \citet{HefTawn2004} residual approach as we get no events as extreme as that observed at Lancaster in $10^8$ events simulated all of which exceed the observed 2015 event at the Cumbrian gauge. In contrast, the Gaussian copula approach gives estimated probabilities which are stable with respect to the conditioning gauge and are computationally efficient in contrast to the existing approach for such an extreme and widespread event.

\section{Discussion}\label{sec:Dis}

Through using semi-parametric model-based inference this paper has shown how the methodology of \citet{HefTawn2004} can be extended to produce more efficient inferences, particularly as the dimension of the multivariate problem increases. Our approach proposed improvements in the inference of the residual distribution of the \citet{HefTawn2004} model; via kernel smoothed-marginal distributions and using a Gaussian copula. These methods also help in terms of computational and statistical efficiency in dealing with the problem of missing data that is commonly encountered in environmental data sets. 



Our proposed Gaussian copula approach has a downside in that a different correlation matrix $\Sigma$ is required for each conditioning site. Thus for $d$ sites there are $d\binom{d-1}{2}$ correlation parameters to estimate, i.e., $O(d^{3})$ parameters. As a result it seems sensible to determine whether there are any known relationships that can help to make the model parsimonious.  An approach suggested by a referee was to adopt a semi-parametric specification method similar to that of \citet{De2014}, whereby the different residuals densities are inter-linked via a tilting term, i.e., 
\vspace{-0.5cm}
\begin{center}
\begin{equation}
\log\left(\frac{g_{i}(\mathbf{z})}{g_{1}(\mathbf{z})}\right)=\gamma_{i}+\mathbf{z}^{T}\delta_{i},~\mbox{for~}i=2,\ldots,d
\label{eq:ExTi}
\end{equation}
\end{center}
with $g_{i}(\mathbf{z})=dG_{i}(\mathbf{z})/d\mathbf{z}$, with $G_i$ the limiting distribution in expression~\eqref{eq:LimPro} when conditioning on variable $Y_{i}$ being large, and with $(\gamma_i, \delta_i)$ being constants. If condition~\eqref{eq:ExTi}, holds the number of parameters reduces to $O(d^{2})$. Unfortunately this formulation does not appear to be appropriate for our residual data either before or after standardisation to Gaussian marginals. An alternative $O(d^{2})$ approach would be to use a stationary Gaussian process to explain $\mathbf{Z}^{N}$ \citep{Tawn2018}, but that requires the process to be modelled in an appropriate space. In standard environmental studies, the Euclidean distance metric between sites is used to explain spatial dependence. However, as shown by \citet{KeefH2009} and \citet{Asadi2015}, Euclidean distance is not always sufficient for capturing the dependence between river flow gauges. The more appropriate distance metric is to consider the hydrological distance, which is defined as the distance between centroids of the associated catchments for each site. This takes into account that two gauges that spatially might be far apart in fact are similar in nature as they lie within the same catchment.	

\begin{figure}[!h]
\centering
\begin{subfigure}[c]{.4\linewidth}
\includegraphics[width=\textwidth]{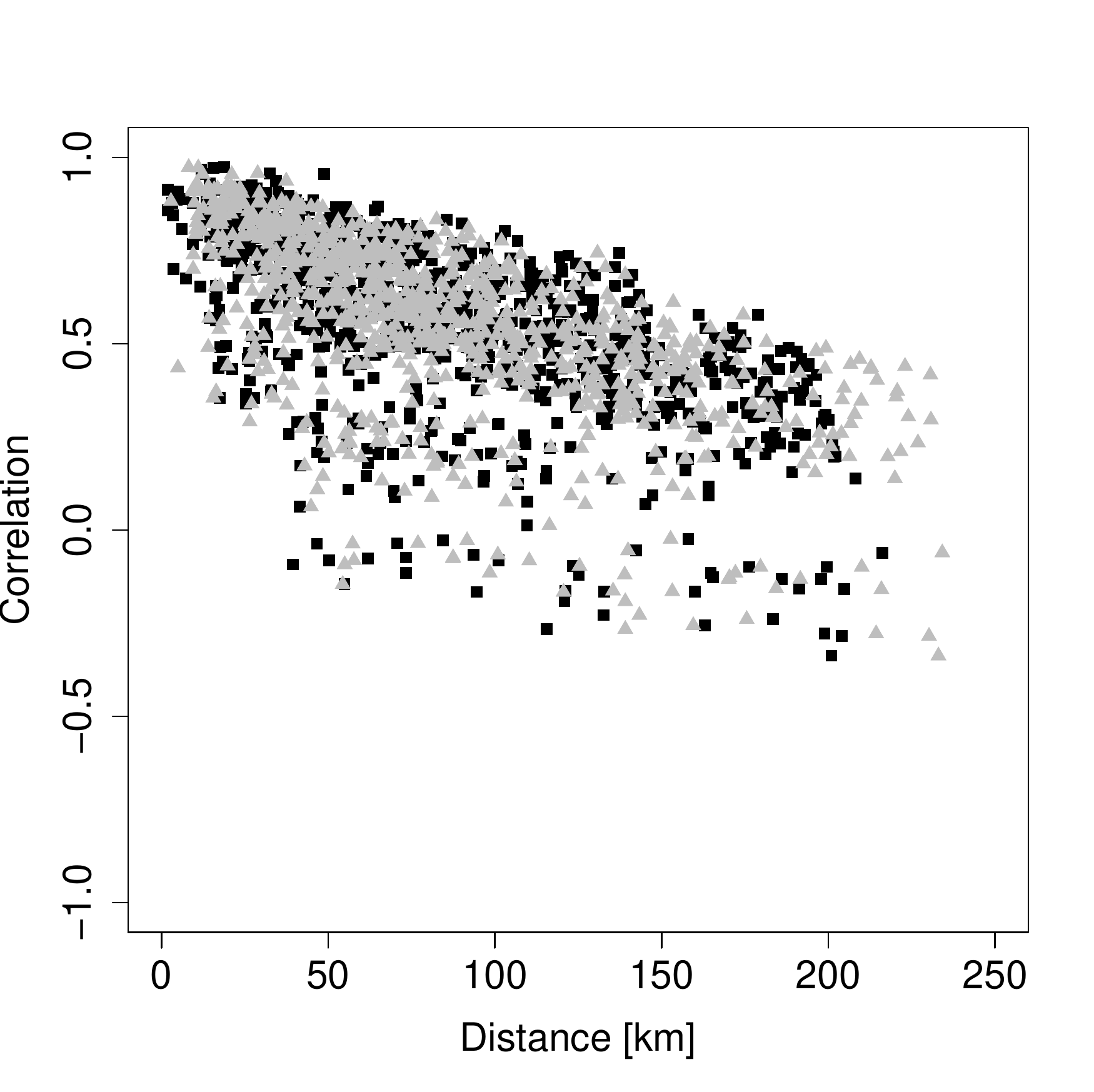}
\caption{Gauge 68003}
\label{fig:G68003}
\end{subfigure}
\begin{subfigure}[c]{.4\linewidth}
\includegraphics[width=\textwidth]{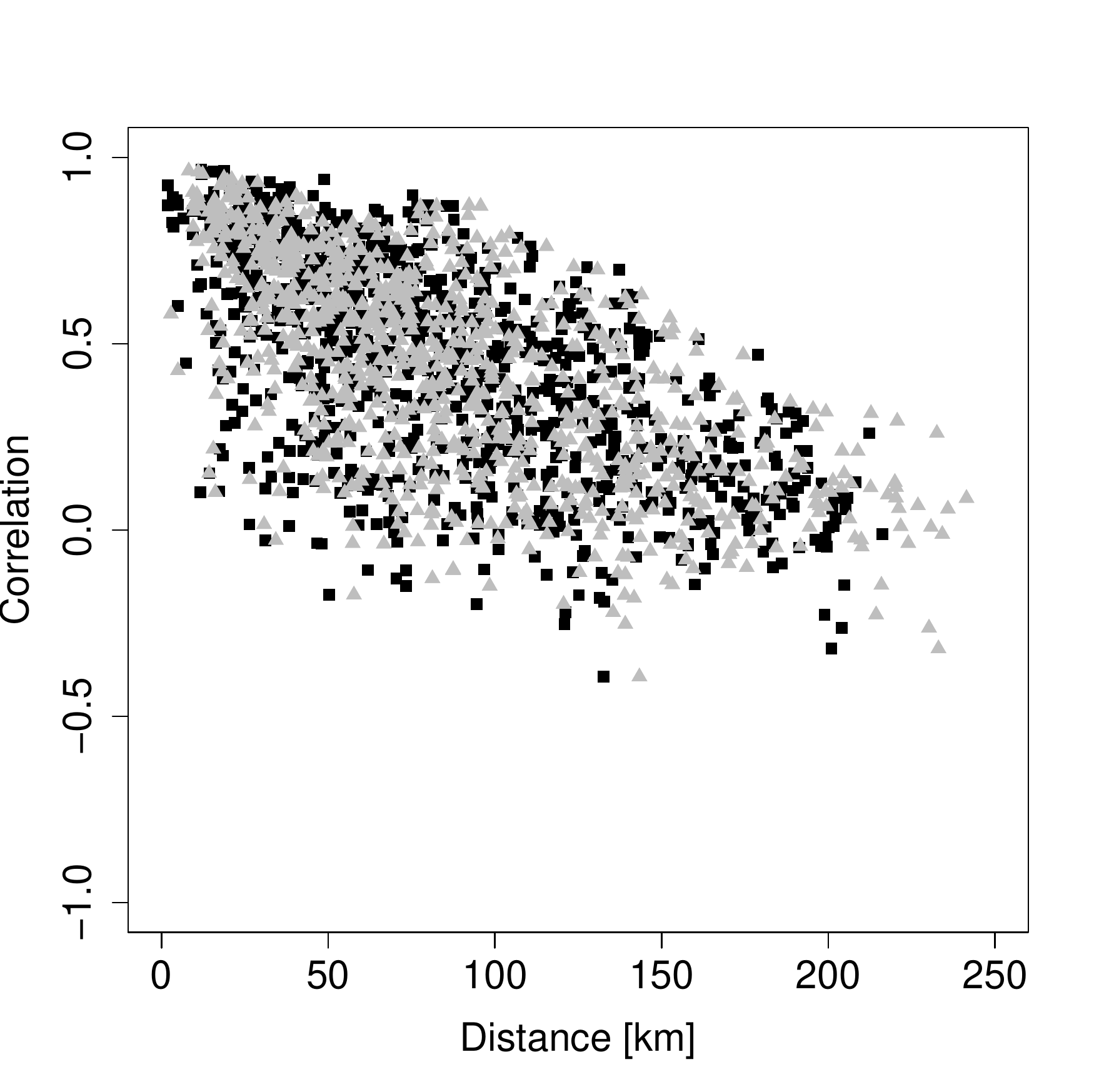}
\caption{Gauge 69017}
\label{fig:G69017}
\end{subfigure}
\begin{subfigure}[c]{.4\linewidth}
\includegraphics[width=\textwidth]{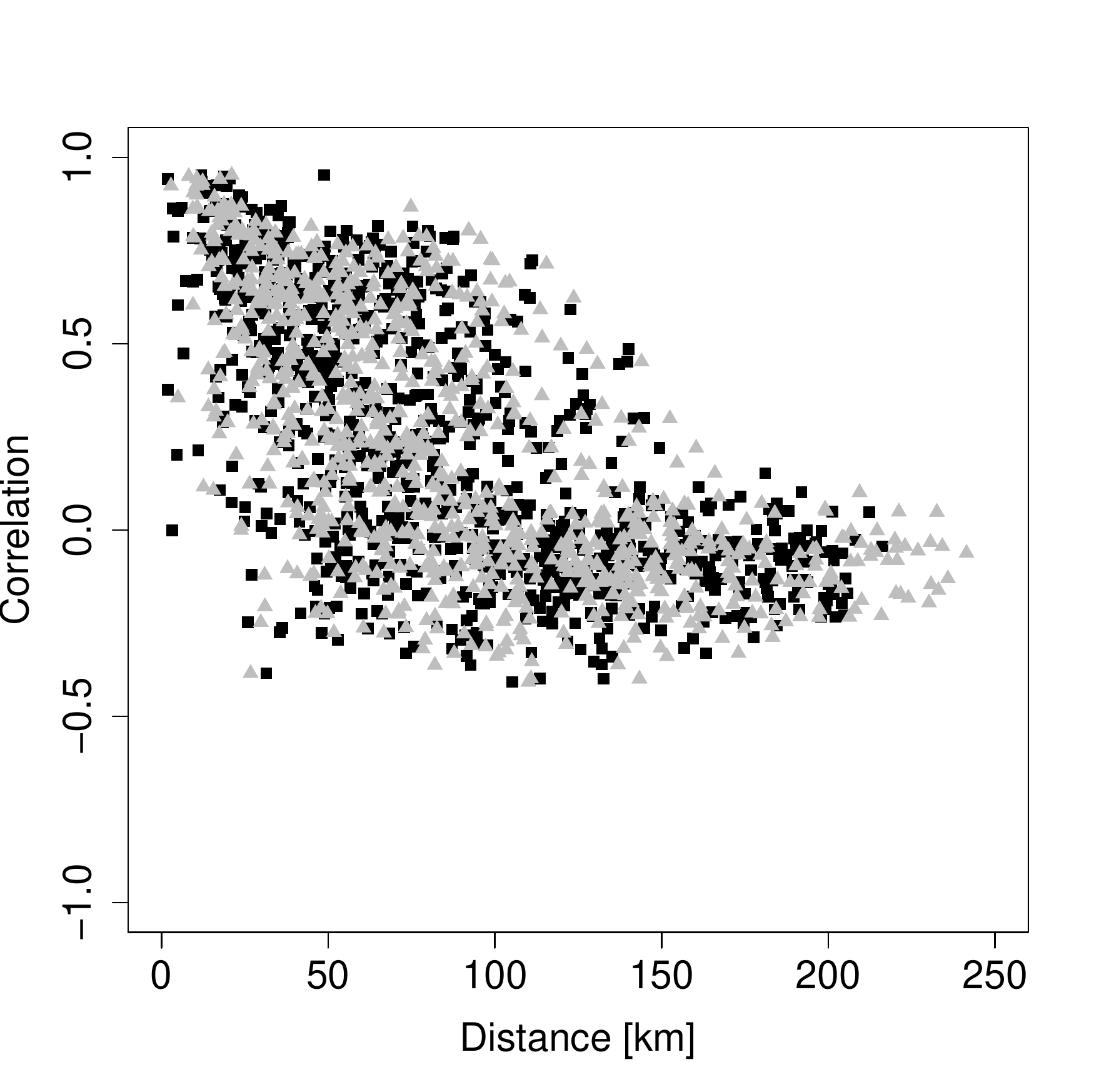}
\caption{Gauge 71001}
\label{fig:G71001}
\end{subfigure}
\begin{subfigure}[c]{.4\linewidth}
\includegraphics[width=\textwidth]{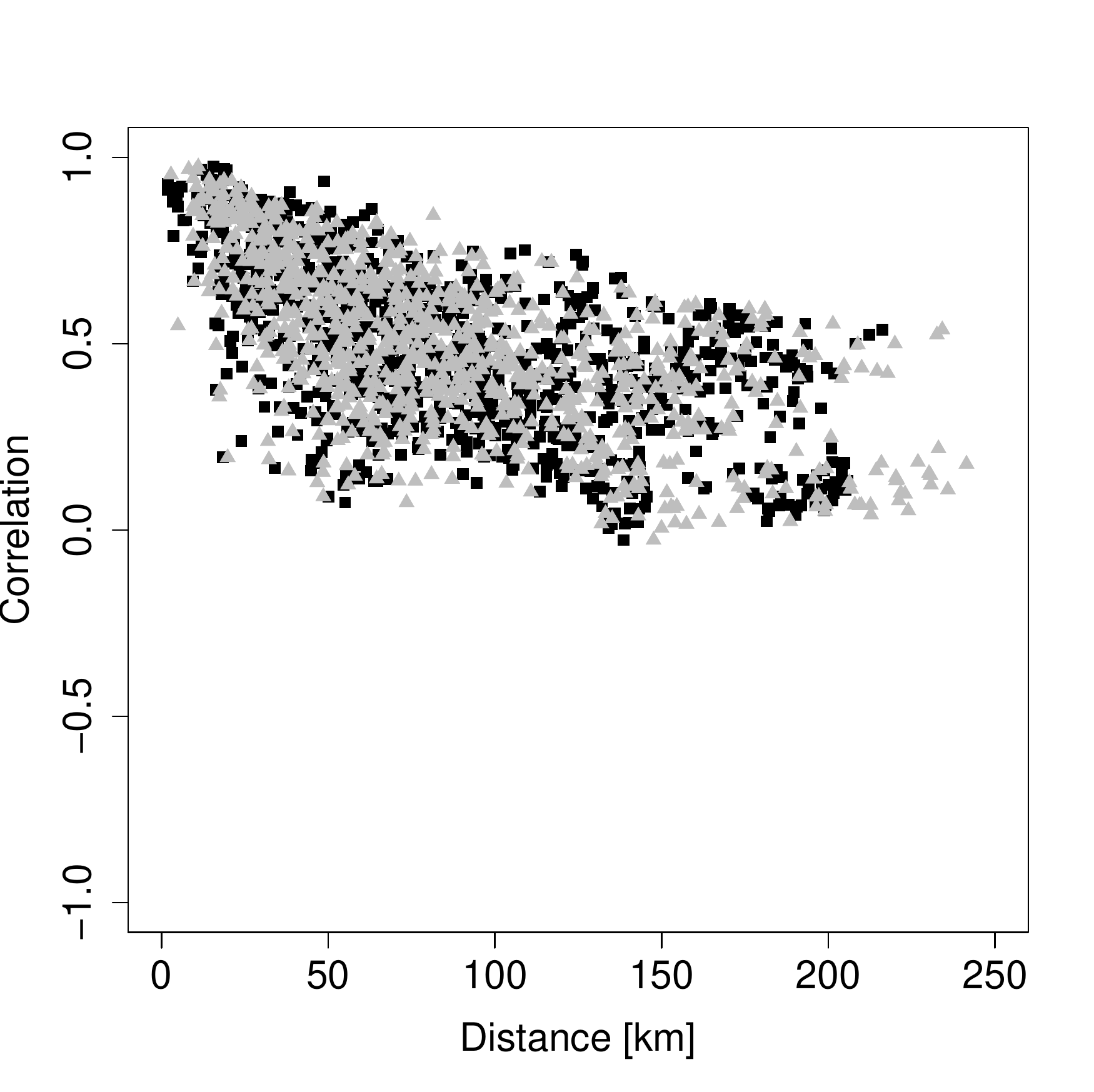}
\caption{Gauge 74001}
\label{fig:G74001}
\end{subfigure}
\caption{The correlation of each pair of residuals against Euclidean (black) and hydrological (grey) distance. Panels four conditioning gauges from the National River Flow Archive: (a) 68003 (south of Manchester), (b) 69017 (western side of the Peak District), (c) 71001 (river Ribble) and (d) 74001 (a small catchment in the Lake District).}
\label{fig:CorMatEucHydro}
\end{figure}

In order to determine whether this factor could be used to simplify the correlation matrix, four conditioning sites were selected with differing spatial locations and catchment areas. Conditional on location $k$, the estimates of correlation between $Z_{i}$ and $Z_{j}$ (for sites $\mathbf{s}_{i}$ and $\mathbf{s}_{j}$) given $Y_{k}$ is large, denoted $\rho_{ij|k}$ for $i,j \neq k$, were plotted as a function of both the Euclidean $||(\mathbf{s}_{i},\mathbf{s}_{j})||_{E}$ and hydrological $||(\mathbf{s}_{i},\mathbf{s}_{j})||_{H}$ distance for each pair. This comparison of the correlation and distance metrics can be seen in Figure~\ref{fig:CorMatEucHydro}. As expected as the distance between pairs of sites increases the correlation tends to decrease. Interestingly, there is no substantial difference between the explanatory capabilities of Euclidean and hydrological distance. Anomalous behaviour can be seen in panel Figure \ref{fig:G68003}, as for one of the sites the residual correlation with all other sites is approximately equal to zero. This site is close to conditioning gauge 68003, therefore the \citet{HefTawn2004} model has explained all of extremal behaviour at this gauge, with the other sites. This illustrates that $\rho_{ij|k}$ will depend on $\mathbf{s}_{k}$ as well.  Other known hydrological characteristics could also be used to explain the residual dependence structure, these include variables such as the catchment responsiveness as well as the soil type. For example, a chalk catchment is slower to respond to heavy rainfall events than a catchment in north west England \citep{Boorman1995}. Generalising these features is difficult as we are trying to simplify the correlation of unexplained behaviour of the extremes rather than of the observed process itself.

The paper has shown that the proposed Gaussian copula model for the joint residual distribution of the \citet{HefTawn2004} model is ideal for classes of asymptotically dependent and asymptotically independent distributions. A simulation study shows in low- and high -dimensional examples the benefits of the proposed approach over other alternatives
for both missing and non-missing data problems as well as under mis-specification of the Gaussian copula. A case study of river flow data shows the benefits of the method for assessing the risk of an event similar to the storm Desmond event. An analogous analysis using existing methods would have been both incredibly computationally expensive and numerically sensitive to the choice of conditioning variable to estimate using existing methods.

\section*{Acknowledgements}
Towe's research was supported by Jeremy Benn Associates Ltd and Innovate UK KTP009454 and EP/P002285/1 (The Role of Digital Technology in Understanding, Mitigating and Adapting to Environmental Change). We thank Ye Liu (HR Wallingford) for helpful discussions. We also thank the referees for their comments and suggestions. The daily mean river flow data were obtained through the National River Flow Archive. 
\nocite{Wickham2016}
\bibliographystyle{apalike}	
\bibliography{References}		
\end{document}